\documentclass[sigconf]{acmart}

\usepackage{tabularx}
\usepackage{booktabs}
\usepackage{makecell}
\usepackage[capitalize, noabbrev]{cleveref}
\usepackage{subcaption}
\usepackage{placeins}
\usepackage{siunitx}
\usepackage{enumitem}
\usepackage{calc}
\usepackage{listings}
\usepackage{float}
\AtBeginDocument{%
  \providecommand\BibTeX{{%
    \normalfont B\kern-0.5em{\scshape i\kern-0.25em b}\kern-0.8em\TeX}}}

\setcopyright{acmlicensed}
\copyrightyear{2025} 
\acmYear{2025} 
\setcopyright{cc}
\setcctype{by}
\acmConference[CHI '25]{CHI Conference on Human Factors in Computing Systems}{April 26-May 1, 2025}{Yokohama, Japan}
\acmBooktitle{CHI Conference on Human Factors in Computing Systems (CHI '25), April 26-May 1, 2025, Yokohama, Japan}\acmDOI{10.1145/3706598.3713554}
\acmISBN{979-8-4007-1394-1/25/04}

\begin{document}

\title{Exploring Mobile Touch Interaction with Large Language Models}

\author{Tim Zindulka}
\email{tim.zindulka@uni-bayreuth.de}
\orcid{0009-0009-1972-351X}
\affiliation{%
  \institution{University of Bayreuth}
  \city{Bayreuth}
  \country{Germany}
}

\author{Jannek Sekowski}
\email{jannek.sekowski@uni-bayreuth.de}
\orcid{0009-0006-3324-7837}
\affiliation{%
  \institution{University of Bayreuth}
  \city{Bayreuth}
  \country{Germany}
}

\author{Florian Lehmann}
\email{florian.lehmann@uni-bayreuth.de}
\orcid{0000-0003-0201-867X}
\affiliation{%
  \institution{University of Bayreuth}
  \city{Bayreuth}
  \country{Germany}
}

\author{Daniel Buschek}
\email{daniel.buschek@uni-bayreuth.de}
\orcid{0000-0002-0013-715X}
\affiliation{%
  \institution{University of Bayreuth}
  \city{Bayreuth}
  \country{Germany}
}

\renewcommand{\shortauthors}{Zindulka et al.}

\definecolor{TimsColor}{rgb}{0.1,0.5,0.8}
\newcommand{\tim}[1]{\textsf{\textbf{\textcolor{TimsColor}{[Tim: #1]}}}}
\definecolor{JanneksColor}{rgb}{0.5,0.8,0.5}
\newcommand{\jannek}[1]{\textsf{\textbf{\textcolor{JanneksColor}{[Sven: #1]}}}}
\definecolor{FlosColor}{rgb}{0.9,0.1,0.8}
\newcommand{\flo}[1]{\textsf{\textbf{\textcolor{FlosColor}{[Flo: #1]}}}}
\definecolor{DanielsColor}{rgb}{0.9,0.6,0.1}
\newcommand{\daniel}[1]{\textsf{\textbf{\textcolor{DanielsColor}{[Daniel: #1]}}}}

\newcommand{\minsec}[2]{\SI{#1}{\minute} \SI{#2}{\second}}
\newcommand{\mins}[1]{\SI{#1}{\minute}}
\newcommand{\secs}[1]{\SI{#1}{\second}}
\newcommand{\pct}[1]{\ifnum\pdfstrcmp{#1}{X}=0
        X\% 
    \else\SI{#1}{\percent}\fi
}
\newcommand{\lbparagraph}[1]{\paragraph{#1}\mbox{}\\}

\newcommand{\lmmci}[5]{$\beta$=#1, SE=#2, CI$_{95\%}$=[#3, #4], p#5}
\newcommand{\posthoc}[2]{#1, p#2}

\definecolor{deemphColor}{rgb}{0.4,0.4,0.4}
\newcommand{\deemph}[1]{\textcolor{deemphColor}{#1}}

\newcommand{\pinch}{pinch-to-shorten}
\newcommand{\Pinch}{Pinch-to-shorten}
\newcommand{\spread}{spread-to-generate}
\newcommand{\Spread}{Spread-to-generate}
\newcommand{\visbubble}{Bubbles}
\newcommand{\visline}{Lines}
\newcommand{\visnone}{NoVis}
\newcommand{\modeours}{Gestures}
\newcommand{\modegpt}{ChatGPT}

\newcommand\revision[1]{\textcolor{black}{#1}}

\begin{abstract}
Interacting with Large Language Models (LLMs) for text editing on mobile devices currently requires users to break out of their writing environment and switch to a conversational AI interface. 
In this paper, we propose to control the LLM via touch gestures performed directly on the text.
We first chart a design space that covers fundamental touch input and text transformations.
In this space, we then concretely explore two control mappings: \spread{} and \pinch{}, with visual feedback loops.
We evaluate this concept in a user study (N=14) that compares three feedback designs: no visualisation, text length indicator, and length + word indicator. 
The results demonstrate that touch-based control of LLMs is both feasible and user-friendly, with the length + word indicator proving most effective for managing text generation. 
This work lays the foundation for further research into gesture-based interaction with LLMs on touch devices.
\end{abstract}



\begin{teaserfigure}
  \centering
  \includegraphics[width=\textwidth]{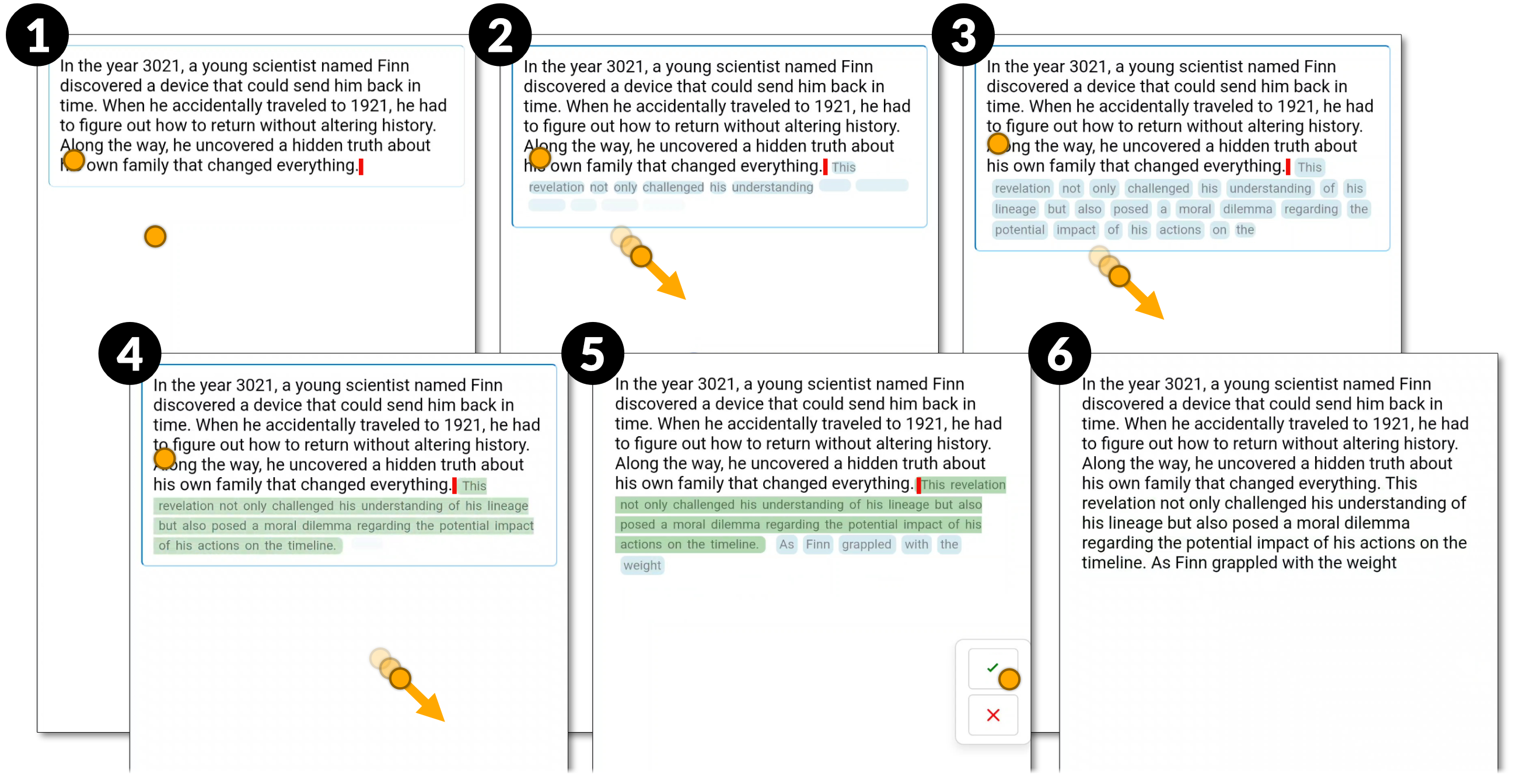}
  \caption{Our \textit{\spread} touch gesture for controlling generative AI on mobile devices. Touches are marked in orange. \textit{(1)} Placing two fingers on the screen sets the cursor (red) to the end of the sentence at the first touch (here: top touch). \textit{(2-3)} Spreading the two fingers fades in blue ``word bubbles'', which indicate estimates of length and number of words to be generated. In the background, an LLM generates text and streams it to the UI, where it is inserted into empty bubbles as it becomes available. \textit{(4)} Reaching the end of a sentence turns the word bubbles into one green sentence bubble. Further spreading the fingers starts generating another sentence. \textit{(5)} A confirmation widget is shown when lifting the fingers. Tapping the check mark accepts the generated text for \textit{(6)} the final result.}
  \label{fig:teaser}
  \Description{This figure shows six stages of the '\spread' gesture for controlling a generative AI system on mobile devices. The gesture is used to generate and insert text into a passage, and key touch points are highlighted in orange circles. Panel 1: The user places two fingers on the screen, and the cursor (shown in red) is positioned at the end of the sentence, where text generation will begin. Panel 2: As the user begins spreading their fingers apart, blue 'word bubbles' start appearing between the cursor and the point of finger contact. These bubbles represent placeholders for the AI-generated words that will be inserted into the sentence. Panel 3: As the spreading gesture continues, more 'word bubbles' are generated, reflecting additional text the AI will insert. The bubbles indicate the length and number of words that are likely to be generated. Panel 4: Once the AI has completed generating text for the current sentence, all the blue 'word bubbles' turn into a single green sentence bubble. This indicates that the AI has fully generated the sentence. Panel 5: A confirmation widget appears in the bottom right corner with a checkmark and 'X' button after the user lifts their fingers. This allows the user to either accept or reject the generated text. Panel 6: The user taps the checkmark to confirm and accept the AI-generated text, which is then finalized and inserted into the document. Each panel shows the progression of the gesture, from starting the text generation process to confirming the final result.}
\end{teaserfigure}

\maketitle

\section{Introduction}
Writing on the go is cumbersome.
Generative AI for text promises writing support, such as with recent features based on the text generation capabilities of Large Language Models (LLMs).
Today, many people interact with LLMs via a conversational user interface (CUI), popularised by OpenAI's ChatGPT application.\footnote{\url{https://chatgpt.com/}}

However, while writing, this introduces the need for a separate application window, besides the text editor. 
On a large desktop screen, users might open both in parallel. Additionally, recent related research as well as industry products\footnote{e.g. Copilot in Microsoft Word: \url{https://copilot.cloud.microsoft/en-US/copilot-word} and Gemini in Google Docs: \url{https://support.google.com/docs/answer/14206696}} have introduced elements like a \textit{sidebar}, to integrate a text editor with a UI for accessing LLMs (e.g. \textit{Wordcraft}~\cite{Yuan20222wordcraft}).

Unfortunately, this solution does not work well with the limited screen space of many mobile devices, such as smartphones. 
As a consequence, mobile users have to engage in cumbersome \textit{context switches}, such as switching between their writing app and a browser or dedicated AI app (e.g. the ChatGPT app\footnote{\url{https://openai.com/chatgpt/download/}}) \revision{to prompt the model}.

\revision{An emerging alternative to writing prompts is direct manipulation.}
Recently, \citet{directGPT} \revision{explored this via mouse and keyboard for composing prompts from textual and visual elements,} including a use case for text editing.
\revision{%
In our work here, we explore direct interaction with LLMs for a new context and goal: using \textit{continuous gestures}, instead of writing prompts, for 
controlling
text generation on \textit{mobile touch devices}, motivated by reducing the need for context switches between separate writing and prompting UIs. %
}

Concretely, we explore this idea with these three guiding research questions:

\begin{enumerate}
    \item What are the fundamental design decisions for mobile touch interaction with generative AI for text?
    \item How might we design for a continuous control loop in this context?
    \item How do users perceive and interact with such a mobile touch interaction technique for generative AI for text?
\end{enumerate}

To address these questions, we first charted a design space that captures a set of core design choices for mobile touch interaction with generative AI for text. 
We then used this space to select and design one fundamental gesture control, \textit{\spread{}} -- plus its counterpart, \textit{\pinch{}}. 
Through a formative study, we identified visual feedback as a key requirement and challenge. We iteratively designed and implemented a solution that also handles irregular latency when streaming generated text from LLMs, using a novel concept of ``word bubbles'' (\cref{fig:teaser}).
Finally, we evaluated our design in a controlled lab study with two parts: Experiment 1 compared the word bubbles against two other alternatives (line highlighting, no visual feedback) in sentence generation/deletion tasks. Experiment 2 compared our gesture concept with a CUI (designed to mimic ChatGPT as an app) in a text editing task.

Our results show that visual feedback improves the control loop in terms of speed, workload, and usability. In particular, interaction with the word bubbles was fastest, reduced overshooting, and was clearly preferred by participants. The gesture interaction also outperformed the CUI baseline.%

\revision{In summary, we contribute: 
(1) The first design space specifically for mobile touch interaction with LLMs, 
to provide structure for exploring, describing, and reflecting on new interaction designs in this emerging context; %
(2) a functional prototype that implements two new gesture controls for LLMs (\spread{}, \pinch{}); %
and (3) insights from a user study evaluation, which show that continuous touch gesture control for LLMs is both feasible and user friendly, with our new ``Bubbles'' feedback design proving most effective for control both in terms of user preference and interaction metrics.
}

In a broader view, this work lays the foundation for gesture-based interaction with generative AI on touch devices and we invite the community to join us in exploring it further.

\section{Related Work}\label{sec:related_work}

\subsection{\revision{Writing and AI}}
\revision{
The integration of AI has become a central topic in recent HCI research on writing assistant systems. 
\citet{Lee2024dsiiwa} recently summarized trends and mapped out a comprehensive design space for this area of work.
Many of their surveyed AI tools exhibit a ``fragmentation'' of text and UIs, as described by \citet{Buschek2024collage}: These tools introduce snippets, cards, pop-ups, and so on, to make space for prompt input and AI output. %
}

\revision{
However, the limited space on mobile devices makes these fragmented interfaces an unsuitable design choice for writing on the go.
Traditionally, mobile text entry research integrated ``intelligent'' features into the keyboard and text (e.g. word suggestions, auto-corrections~\cite{Palin2019, Banovic2019, Quinn2016}). However, today, LLMs are capable of much larger text contributions, beyond individual words or short phrases that had found their place at the top of mobile keyboards. 
Notably, none of the 115 tools surveyed by \citet{Lee2024dsiiwa} is explicitly designed for mobile touch use (beyond keyboard apps), highlighting a gap in research and practical applicability for this context.}

\revision{At the outset of our research, this motivated us to first chart a novel design space for mobile touch interaction with LLMs. We then used this space to inform our new design of gesture-controlled text generation specifically tailored for writing on mobile devices. Concretely, our resulting design supports users in triggering text generation, anticipating its length, and potentially curating it (cutting, regenerating) -- in one continuous interaction via familiar touch gestures (spread, pinch). At the same time, it supports users in handling the irregular system latency of text generation with a new visual control loop design (feedforward/feedback). %
}

\subsection{Beyond Prompts and Towards Continuous Interaction}
With \textit{DirectGPT}, \citet{directGPT} recently explored direct manipulation paradigms for interacting with LLMs, fundamentally inspiring our work. 
We extend this to touch gestures and mobile devices. \revision{In the future, we see} the potential for combining these concepts, such as integrating our gestures for text generation with the drag-and-drop interactions by \citet{directGPT} to refer to objects beyond text.
\revision{\textit{DirectGPT} operates in a discrete, turn-based interaction model: While the user's direct manipulation actions might be continuous (e.g. drag and drop), they ultimately result in a prompt that is ``done and ready'' to send to the LLM. The user then evaluates the resulting output to inform their next prompting actions.}
Similarly, \textit{TaleBrush} by \citet{Chung2022taleBrush} lets users steer story character narratives via drawn lines, using continuous (mouse) input. It also generates text only after input completion.
\revision{In contrast, our approach aims for continuous direct interaction, where continuous user actions (e.g. spreading fingers) are automatically and continuously sent to the LLM in the background, and output is continuously updated in the UI in parallel, approaching a new ``closed'' direct manipulation control loop.}

Current systems like ChatGPT already use continuous displays for output, that is, adding one word at a time, to simulate an impression of ``writing'' and to handle computational latency in the UI.
Building on this, \citet{Lehmann2022suggVsCont} evaluated a design where users intervene in an LLM's continuous output stream for co-writing on smartphones. This pushed users into an editing role, requiring them to delete and modify text to align it with their intentions.

This role distribution -- AI drafts, user edits -- may suit cases where users prefer avoiding manual typing to generate initial drafts. Mobile devices, where text entry can be cumbersome~\cite{Kristensson2014inviscid}, are a prime example. 
However, research shows that AI-generated text often fails to meet user expectations, particularly in interpersonal communication~\cite{Fu2024texttoself, Liu2022aimailperception, Robertson2021cantreply}, and tends to be verbose~\cite{Fu2024texttoself}. These limitations highlight that while adding text to drafts is relatively simple, interacting with generated text is far more challenging, despite clear motivations for doing so.
\revision{We explore continuous direct interactions to address these challenges in the space-constrained context of mobile touch devices.} %

\subsection{Designing Mobile Touch Gesture Controls}
One \revision{existing} gesture-based UI for mobile text entry is the word-gesture keyboard~\cite{Zhai2012gesturekbMagazine}. It has inspired gestures for formatting~\cite{Alvina2017commandboard}, and can be combined with gestures for triggering word corrections~\cite{Cui2020justcorrect, Zhang2019typthencorrect}. \revision{These gestures are typically visualised via finger trajectories. More broadly, visual feedback} is a key component of designing gesture controls.

\subsubsection{Feedback via Additional Visual Elements}
The well-known \textit{marking menus}~\cite{Kurtenbach1993markingmenus} \revision{draw} the pointer trajectory as a line, adding a line segment per submenu the user \revision{moves} through. This guides learning of gesture mappings and can then be ignored by expert users~\cite{Kurtenbach1994markingmenus}. 
With \textit{Fieldward} and \textit{Pathward}, \citet{Malloch2017fieldpathward} designed visual feedback that is continuously updated while moving the finger, to guide users in defining a distinguishable set of gestures. Many gestures in our context of interacting with an LLM are also likely to benefit from visual feedback that is updated ``live'', to create a closed control loop.
Concrete requirements for visual feedback differ in our case: For example, visualising finger trajectories may not be suitable, since it might occlude the text the user is interacting with. 
Occlusion has been addressed in concepts for mobile text selection, such as a pop-up (callout) displayed above the finger position~\cite{Ishii2016callout, Vogel2007shift}. This is designed for focusing on the word or character at the cursor. In contrast, many LLM-based applications are likely \revision{interested in larger scopes of text} (e.g. generating or revising a paragraph or whole draft).

\subsubsection{Feedback without Additional Visual Elements}
Other established mobile touch gestures do not need additional visual markup as feedback, because their triggered transformations are already visual in nature. Examples include scrolling by swiping with the finger vertically, and zooming in and out of a text or image via spreading and pinching fingers. The resulting change in content size and/or location \textit{is} the feedback. \revision{In contrast,} we are not interested in geometric zooming or offsets, but in gestures that change the text content in some way.

\subsubsection{Summary and Implications for our Research}
In summary, related literature and industry standards present two kinds of design for gesture control loops on mobile devices -- those that add extra visual elements and those that do not. 
Notably, text is usually in the latter category or text-related visual feedback elements are local in nature. However, we require a novel combination here -- feedback for continuously manipulating (larger scope) text content with a gesture. \revision{We next describe this open design space in more detail.}

\section{A Design Space for Mobile Touch Interaction with LLMs}
\label{sec:design_space}

\begin{figure*}
    \centering
    \includegraphics[width=1\linewidth]{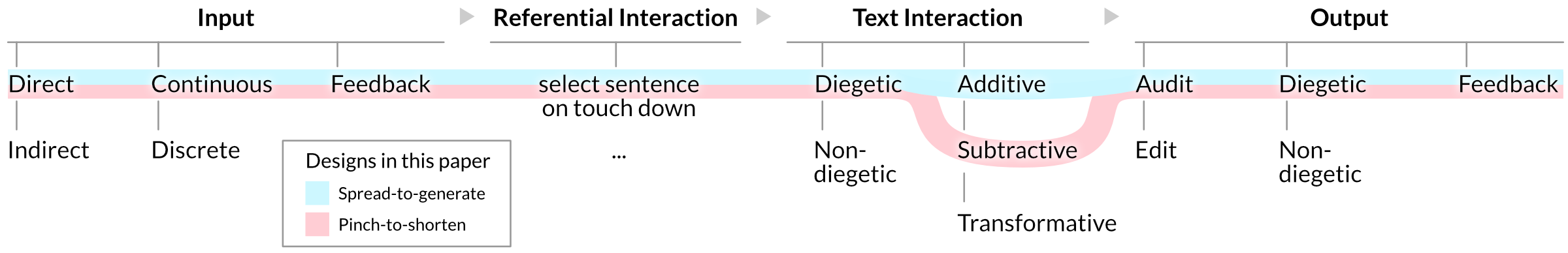}
    \caption{Overview of our design space for mobile touch interaction with generative AI for text, with four dimensions (left to right) and subdimensions (values at vertical lines). Coloured streams indicate the design choices for the two concrete touch gesture controls that we designed, implemented, and evaluated in this paper.}
    \Description{This figure represents a design space for mobile touch interaction with generative AI for text. It spans four dimensions: Input, Referential Interaction, Text Interaction, and Output, each further divided into subdimensions. Coloured streams (blue for ``\Spread{}'' and red for ``\Pinch{}'') highlight the specific design choices evaluated in the paper. Both gestures have the following choices highlighted: ``direct'', ``continuous'', and ``feedback'' for the ``input'' dimension, ``select sentence on touch down'' for the ``referential interaction'' dimension, ``diegetic'' and ``additive'' is marked for spread in the ``text interaction'' dimension whereas pinch has ``diegetic'' and ``subtractive'' highlighted. In the output dimension both have ``audit'', ``diegetic'', and ``feedback marked''.}
    \label{fig:design_space}
\end{figure*}

We developed a design space for touch-based LLM interaction through focused brainstorming sessions and affinity diagramming \cite{hartson2012ux} in our research team. We gathered high-level ideas, examples from research and industry, and explored various mappings of LLM capabilities to touch gestures. 
\revision{The resulting space} has four dimensions (\cref{fig:design_space}):
(1) Input,
(2) Referential Interaction,
(3) Textual Interaction,
(4) Output.

\subsection{Input}
This dimension captures three fundamental choices for interaction designs with touch input. %

\subsubsection{Direct vs Indirect}
Indirect input for interaction with an LLM requires additional UI elements (e.g. virtual keyboard, floating objects, buttons) and/or uses an external application beyond the main writing interface, such as when copy-pasting in text generated with ChatGPT.
In contrast, direct input refers to touch gestures directly on the text that the LLM should operate on. Mobile device users are used to \revision{typical gestures such as} tap, swipe, pinch, spread, and drag. 

\subsubsection{Discrete vs Continuous}
The most common type of touch input today is a discrete event -- tapping a button. Beyond this, there are touch gestures that unfold continuously on the screen over time, such as a swipe or pinch.

\subsubsection{Feedback} 
This captures if and how feedback is given to the user leading up to the final input and/or after it has been completed. For example, this might include a visual trajectory line while performing a gesture. As another example, text selection feedback can serve in this role, such as when highlighting the current selection with a coloured background, updated ``live'' while swiping over it.

\subsection{Referential Interaction} 
Inspired by work on referential and verbal acts in HCI~\cite{wolff1998acting}, we separate these aspects in our design space. Our referential dimension captures \textit{how users interact with text to set the context}, scope, reference, and so on, as a precursor to LLM operations on that text. Visual (UI) changes resulting from these interactions are typically ephemeral and not meant to become a lasting part of the text document.
\revision{As a common example, text selection} may involve (long-)taps, moving the finger over the text, and/or dragging UI instruments~\cite{BeaudouinLafon2000instrumental} (e.g. start/end markers). We decided to keep this dimension at a high level, as it is not our main focus. Possible subdimensions to explore in the future include the design of selection procedures and markers \revision{for} ``pointing'' an LLM at text parts (cf.~\cite{directGPT}).

\subsection{Textual Interaction}
This captures \textit{how users interact with the content (semantics) and style of the text}. These interactions are performed to trigger and parametrise the LLM to operate on the text in a lasting way.
Examples from research on AI writing support~\cite{Lee2024dsiiwa} include extending or shortening text, summarising parts of it, and rewriting, possibly with a specific \revision{goal} (e.g. adjusting tone or level of formality).
To structure this space of possible \revision{AI} text operations, here we focus on two aspects we identified as most fundamental \revision{in} our context.

\subsubsection{Diegetic vs Non-diegetic Text}
This captures whether text involved in interaction (e.g. text selected with a preceding referential interaction) is \textit{diegetic}, \textit{non-diegetic}, or a mix of both (cf.~\cite{Buschek2024collage, Dang2023choice}). Diegetic text \revision{is intended to be in} the final text (e.g. a sentence in a story to improve with AI), while non-diegetic text is not (e.g. a ``todo'' note or prompt for the LLM).

\subsubsection{Additive vs Subtractive vs Transformative}
This subdimension captures whether the AI text operation involved in the interaction is \textit{additive}, \textit{subtractive}, \textit{transformative} or a mix. Additive operations add words, such as when extending a draft. Subtractive refers to the opposite, such as when the system removes words to shorten a text semantically (e.g. by summarising parts of it) or mechanically (e.g. cutting the last 3 sentences). Finally, transformative refers to AI operations that change text without clearly adding or removing \revision{words}. For example, a user might use AI to revise their draft with the goal of achieving a consistently formal tone. This might involve both cutting informal phrases as well as adding parts to adhere to formal expectations.

\subsection{Output}
This dimension captures how the results of the interaction -- the result of the AI's text operation -- is presented.

\subsubsection{Audit vs Edit} 
\revision{Text changes can be} direct edits by the AI (e.g. fixing spelling mistakes) or ``audits'', typically suggestions for users to accept or reject (cf. \cite{Cooper2014aboutface}). 

\subsubsection{Diegetic vs Non-diegetic Text}
AI text resulting from the interaction \revision{can be} diegetic or non-diegetic, or a mix of both. For example, an added sentence suggestion is diegetic, as it is intended to become a part of the text. In contrast, an added feedback comment is non-diegetic AI output (cf.~\cite{Buschek2024collage}).

\subsubsection{Feedback} 
This subdimension captures if and how feedback is given on the way to the final result of an AI text operation and/or after it has been completed. For example, a loading indicator \revision{might show that the} LLM is computing the output. Text itself might be used \revision{for this}, such as in ChatGPT, which adds one word at a time.

\subsection{Applying the design space: \Spread{}, \Pinch{}}
\label{sec:ds_applied}

In this paper, we explore two concrete designs arising from our design space. \cref{sec:implementation} describes \revision{their implementation.} Here, we situate them in our design space to illustrate how the space can be used to generate designs and systematically describe them. %

\subsubsection{\Spread{}}
This touch interaction uses \textit{direct} and \textit{continuous} input (spread gesture with two fingers, as known from zooming) on \textit{diegetic} text (user's draft) to control an \textit{additive} LLM capability with \textit{diegetic} output (text generation) that users need to confirm (\textit{audit}). 
The \textit{referential interaction} is realised by interpreting the first touch event of the two fingers as a selection of the sentence, after which the generated text should be added.
The \textit{feedback} for both input and output is combined into one visual element -- an indicator of the length of the generated text, extended ``live'' while spreading the fingers. \revision{Our study (\cref{sec:method}) compares} two variants of this design (lines, bubbles) with a baseline (no feedback).

\subsubsection{\Pinch{}}
This interaction is the complement to the above: It uses \textit{direct} and \textit{continuous} input (pinch gesture with two fingers) on \textit{diegetic} text (user's draft) to control a \textit{subtractive} operation with a \textit{diegetic} result (text deletion) that users need to confirm (\textit{audit}). 
\textit{Referential interaction} and \textit{feedback} are realised as described above. However, our visual design differs to make it easier for users to perceive the effect at a glance (e.g. red for delete, blue for extend). Our implementation of \pinch{} does not require an LLM.

\section{Concept and Implementation}
\label{sec:implementation}
We propose the concept of controlling LLMs for text operations on mobile devices via touch gestures.
To explore this, we developed a prototype, as described next. %

\subsection{Concept Development}
\revision{Here, we describe our concept development.}

\subsubsection{User Feedback and Design Iteration}
\begin{figure*}[t]
     \centering
     \begin{subfigure}[b]{0.475\textwidth}
        \centering
        \includegraphics[width=0.8\linewidth]{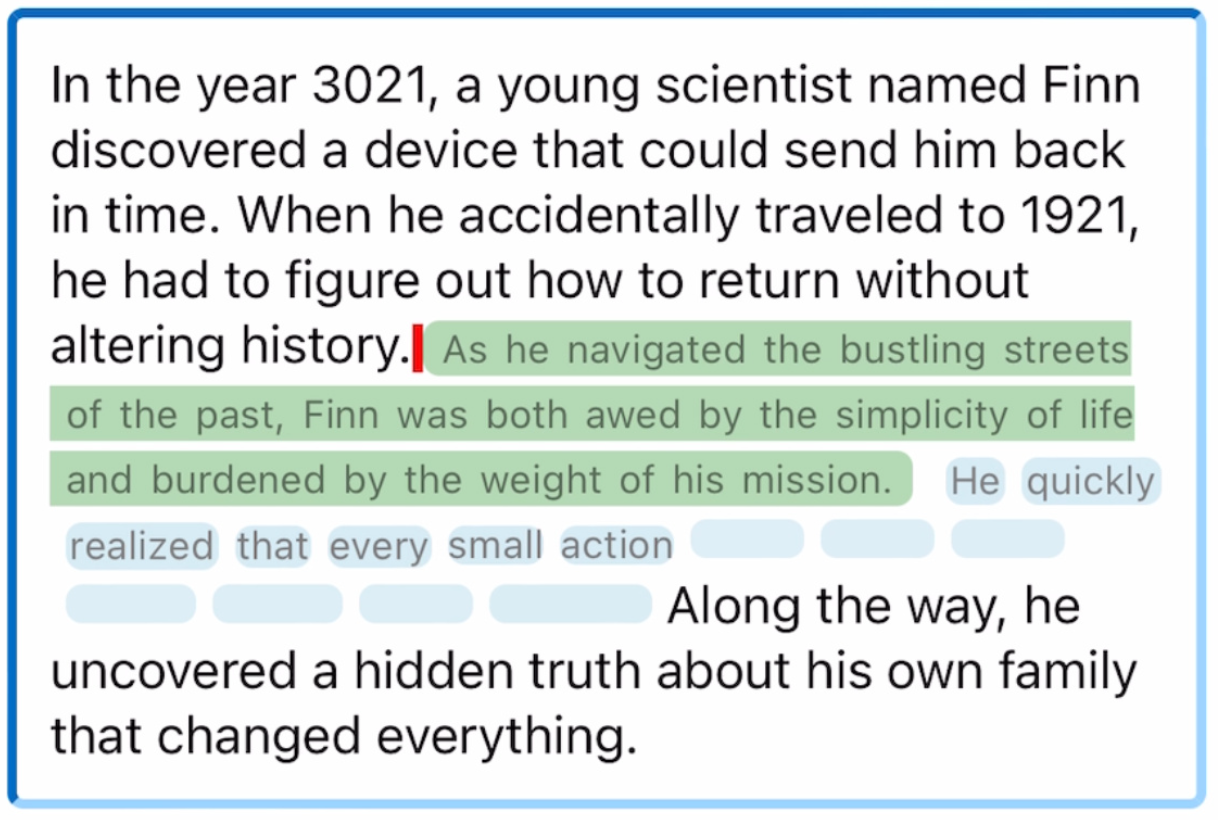}
        \caption{\Spread{} gesture in the editor: the red cursor marks the starting point for generation. Blue bubbles represent the expanding text length, filling with words as they arrive from the backend. Once a full sentence is generated, the individual word bubbles merge into a large green bubble, indicating the completion of the sentence.}
        \label{fig:add_bubbles}
     \end{subfigure}
     \hfill
     \begin{subfigure}[b]{0.475\textwidth}
        \centering
        \includegraphics[width=0.8\linewidth]{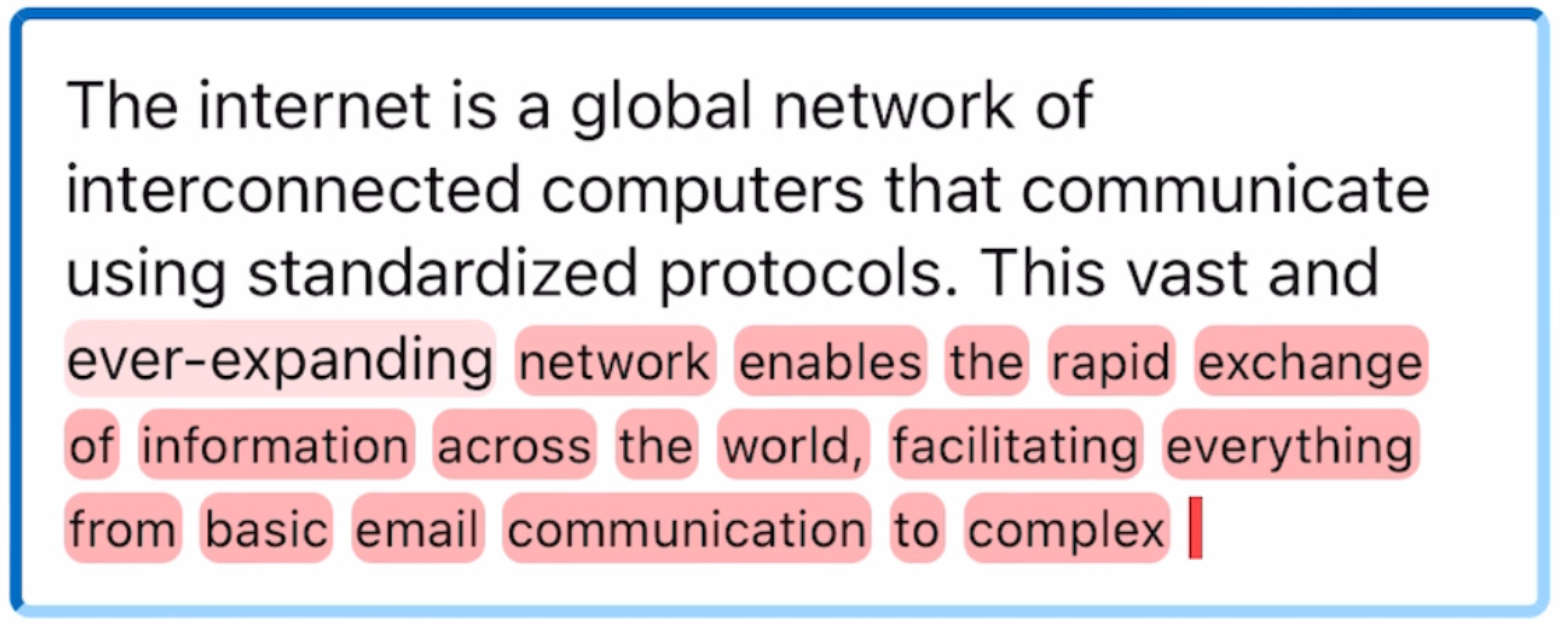}
        \caption{\Pinch{} gesture in the editor: a red cursor marks the starting point for word removal. Dark red bubbles highlight words already selected for deletion, the lighter (and slightly larger) red bubble captures a transitional word animation from unselected to selected.}
        \label{fig:remove_bubbles}
     \end{subfigure}
     \caption{Our frontend enables gesture interaction with text on mobile devices. We employ our Bubbles visual feedback design to communicate essential information to the user.}
     \Description{This figure contains two panels that show different types of gestures -- '\spread{}' and '\pinch{}' -- used to interact with text in a mobile text editor. Each panel illustrates how 'Bubbles' visual feedback helps users understand the effects of these gestures. Panel (a): Depicts the '\spread{}' gesture. A passage of text is shown with a red cursor marking the point where text generation will begin. As the user spreads their fingers, blue word bubbles appear in the text to represent the words being generated in real-time by the AI. Once the user completes a full sentence, the individual blue word bubbles merge into one large green bubble, signifying the end of the sentence. The selected text, shown in green, has been generated during the gesture. This panel highlights the expanding nature of text as the gesture progresses. Panel (b): Illustrates the '\pinch{}' gesture. In this text passage, the red cursor again marks the starting point for the interaction, but in this case, words are being removed. Dark red bubbles appear around the words to be deleted as the user pinches their fingers together. Lighter red bubbles indicate transitional stages of word selection for removal. The dark red bubbles fully encompass words marked for deletion, making it clear which text will be removed.}
     \label{fig:bubbles}
\end{figure*}

Our prototype UI consists of a text field with spread and pinch gestures that trigger text extensions or shortenings based on finger distance. It evolved iteratively with user feedback from a formative study (N=17):  

\begin{itemize}
\item \textit{Cursor and Sentence Selection:} The initial prototype featured no cursor, and text was generated or deleted at the end of a paragraph. As users wished for more control, we added functionality to select a sentence with the first touch of a gesture, placing a cursor at its end (\cref{fig:bubbles}).
\item \textit{Text-Length Indicator:} The lack of feedback during gesture execution caused confusion due to the short delay in text generation. This motivated us to explore visual feedback designs, eventually leading to the \visbubble{} design (\cref{fig:bubbles}). %
\item \textit{Building Structure:} To address users feeling overwhelmed by rapidly extending text, we added structure by visually separating each fully generated sentence and surrounding individual words in incomplete sentences with bubbles.
\item \textit{Displaying the Token Stream:} Users preferred to see generated text immediately, rather than estimating its length first and being surprised by the content. In response, we began streaming the incoming tokens directly into the visual feedback (e.g. bubbles) as soon as they were generated.
\end{itemize}

\subsubsection{Designing a (Visual) Control Loop}
\label{sec:vis_design}
Direct touch interaction with LLMs requires novel feedback designs (\cref{sec:related_work}). 
We designed a control loop for \spread{} and \pinch{}. %
Here we list our \textit{design challenges and goals}, extracted from the literature and our formative study:
\begin{enumerate} 
\item \textit{Presenting All Relevant Information:}
To avoid context switching, all necessary information (e.g., input, output, system state) should be embedded in the text display. Ideally, no additional UI elements, like sidebars or chat boxes, are \revision{needed}.
\item \textit{Managing Generation Latency and Token Streaming:} AI text generation introduces continuous token streams with latency. While less critical in chatbot \revision{UIs, this needs to be handled well for} continuous interaction: \revision{For example, latency may} slow down the experience, while real-time generation may overwhelm users if content appears too fast. %
\item \textit{Integration with Mobile Touch Gestures:} %
\revision{The design needs to handle} these LLM-specific issues in a visual control loop that meets the demands of mobile touch interaction.
\end{enumerate}

We outline four design goals to address these challenges and explain how our implementation meets each of them:
\begin{enumerate} 
\item \textit{Input Visibility:}
The user's input to the LLM for text extension should always be clear. %
\revision{Besides the textual context, this also includes the user's desired amount of text to be generated.}
Our feedback design (\cref{fig:bubbles}) uses colour to \revision{indicate this and to} separate newly generated text from existing text.
\item \textit{Output Clarity:}
Generated/removed text should be visible immediately \revision{when available from the LLM}. %
Our visualisation fills the text-length indicator (e.g. bubbles) ``live''. To \revision{further} enhance clarity, the blue word bubbles merge into a green sentence bubble upon completing a sentence (\cref{fig:teaser}). %

\item \textit{System State Communication:}  
The system’s current state -- processing, generating, or encountering latency -- should be clearly communicated. 
Visual cues must indicate when it is ready for input or generating output. We support this by animating all state changes (e.g. fading in bubbles). %

\item \textit{Integration with Mobile Writing Workflows:}
The system should integrate seamlessly into existing \revision{mobile} writing workflows.
All interactions should occur within the text \revision{UI}, with no context switches.
\revision{Our feedback loop with our \visbubble{} design} allows users to make real-time adjustments as they engage with the LLM \revision{through} familiar spread and pinch gestures.
As UI objects, the bubbles can support further interaction possibilities. While not in our focus, we explore some ideas in our prototype (\cref{sec:additional_features}).
\end{enumerate}

\subsection{Backend Implementation}\label{sec:backend}
The backend \revision{connects} frontend and OpenAI API\footnote{\url{https://platform.openai.com/}} (gpt-4o-mini).
It %
was implemented as a Node.js/Express application. 
We leverage the ``/aiCompletionStream'' API endpoint which forwards each incremental delta received from OpenAI directly to the frontend.
\revision{This setup achieved a mean latency of 242 ms (median: 98 ms, std: 262 ms) from gesture initiation to word display, during our study.}
\cref{sec:appendix_prompts} provides details on our prompting templates.

\subsection{Frontend}
The frontend (\cref{fig:bubbles}) is a web application in React.

\subsubsection{Selecting the Sentence}
\revision{When the first finger of a pinch/spread gesture touches the screen, we find the nearest sentence with a spiral scanning algorithm (cf. \cite{photonics11060540}), which measures touch-sentence distances around} the touch point in an outward spiral pattern.
\revision{Upon detecting the second finger, the system places a red cursor at the end of this nearest sentence. This end is either punctuation (``.'', ``!'', ``?'') or the last word.}

\subsubsection{Mapping Movement to Words}
\label{sec:dist_word_mapping}
The system translates \revision{the user's finger movements} into \revision{an internal} word count, \revision{increasing or decreasing it} for positive or negative distance \revision{changes between the two fingers}, respectively. 
Based on our usability testing, \revision{we map} 1.75\,mm \revision{of distance change to generating or removing one} word. Note that the ``optimum'' might be \revision{device-}specific. %

\subsubsection{Triggering Changes}
When the user lifts one or both fingers, the gesture ends, and the red cursor disappears.
A widget appears at the right \revision{screen} edge (\cref{fig:teaser}.5), \revision{with buttons to confirm or} reject the changes. 
\revision{Confirming a spread integrates the generated text, while confirming a pinch deletes the marked words.} 
Rejecting reverts all changes. 

Alternatively, \revision{the user can resume} the gesture by placing two fingers back on the screen, which makes the cursor reappear and the gesture continue \revision{from where it was left off}. 

\subsubsection{Text-length Indicators}\label{sec:impl_text_length_ind}
We introduce \textit{\visbubble} (\cref{fig:bubbles}) as a feedback design that indicates text-length changes in the control loop for touch-based text generation, addressing the three challenges detailed in \cref{sec:vis_design}.

This design also helps with managing the irregular latency of continuous text generation with LLMs by acting as placeholders, which are filled with words as they become available.
\revision{Concretely, while the fingers move,} the expected word count is estimated based on finger distance changes (see \cref{sec:dist_word_mapping}).
Starting at the cursor position, \revision{the system} adds bubbles one by one until \revision{reaching the current} word count \revision{value}. %
As \revision{actual} generated words become available from the LLM, they are inserted into the bubbles (\cref{sec:impl_text_generation}).

\revision{Thus,} bubbles \revision{are} placeholders for generating text with the spread gesture. \revision{These bubbles} are \textit{blue} (\cref{fig:add_bubbles}).
Once a full sentence is generated, they merge into one \textit{green} bubble. 
\revision{In contrast}, \textit{red} bubbles encapsulate words to be removed by pinching.

The width of each \textit{empty} blue bubble (placeholder) is determined by a function that simulates (randomly) varying word lengths. %
For each, we roll a length between 5 and 10 characters as assumed typical word lengths, and assume five pixels per character. 
Finally, when a placeholder bubble \revision{appears, it blinks shortly} to indicate that a word will appear. %

\subsubsection{Text Generation}\label{sec:impl_text_generation}
\revision{During the gesture, we use a buffer:} Text generation is triggered only if the buffer of already generated words is insufficient to fill all empty bubbles. \revision{If so,} a system prompt (see \cref{prompt:extend}) is sent to the backend together with the text  \revision{preceding} the cursor, plus all words generated up to this point.
The backend streams the response in chunks, allowing placeholders to be filled in real-time without waiting for the entire response.

\revision{In detail,} streamed chunks are processed by handling incomplete words, filtering and combining special characters, updating buffers, and filling bubbles. %
In pinch mode, empty bubbles are removed, text-filled ones push their content back to the buffer, end-of-sentence bubbles lose their last word, and bubbles in incomplete sentences remove the last word without colour change, ensuring smooth feedback.
As a result, if users are unsatisfied with a generated sentence, they can pinch the sentence to remove it and spread again to regenerate a new sentence.

\subsection{Additional Features}\label{sec:additional_features}
We explored extensions to our core functionality that aim to improve usability and extend interaction possibilities.

\begin{figure*}[t]
     \centering
     \begin{subfigure}[b]{0.475\textwidth}
        \centering
        \includegraphics[width=0.9\linewidth]{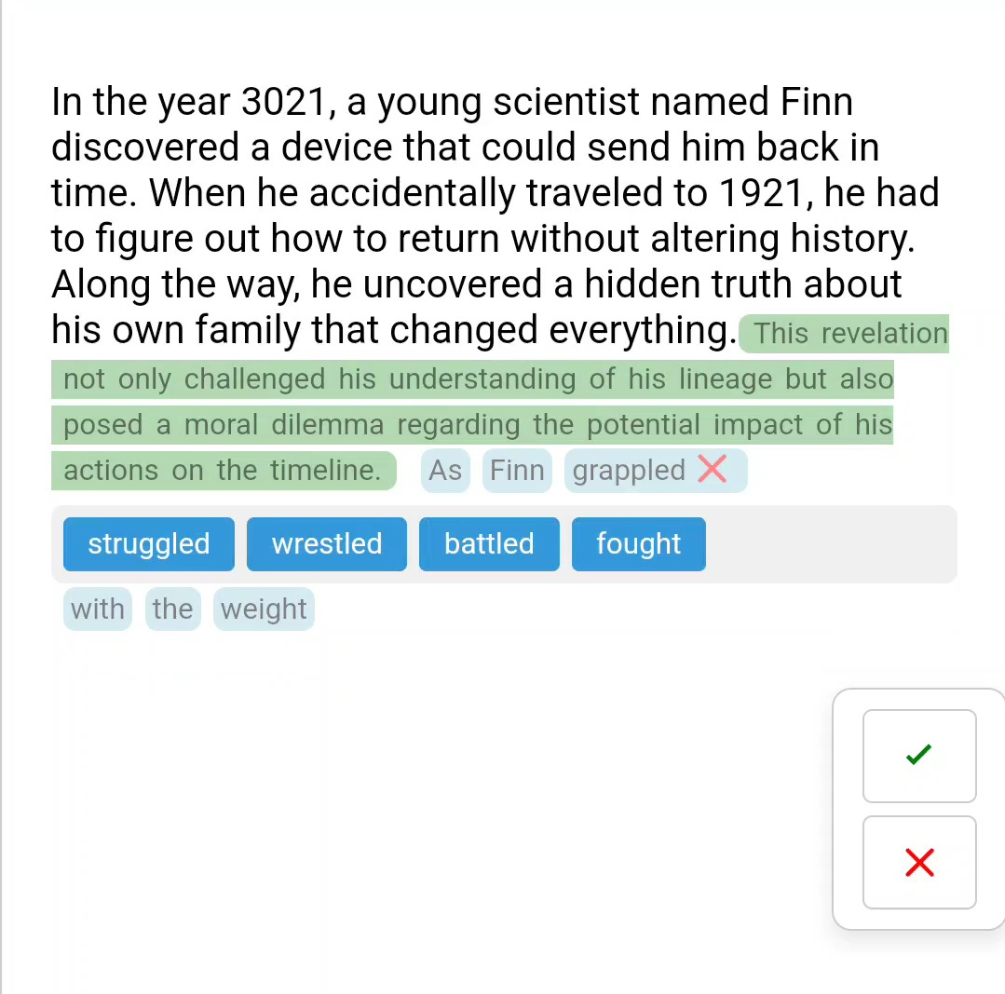}
        \caption{When performing a longpress on a word bubble, suggestions appear beneath it.\newline}
        \label{fig:long_press_words}
     \end{subfigure}
     \hfill
     \begin{subfigure}[b]{0.475\textwidth}
        \centering
        \includegraphics[width=0.82\linewidth]{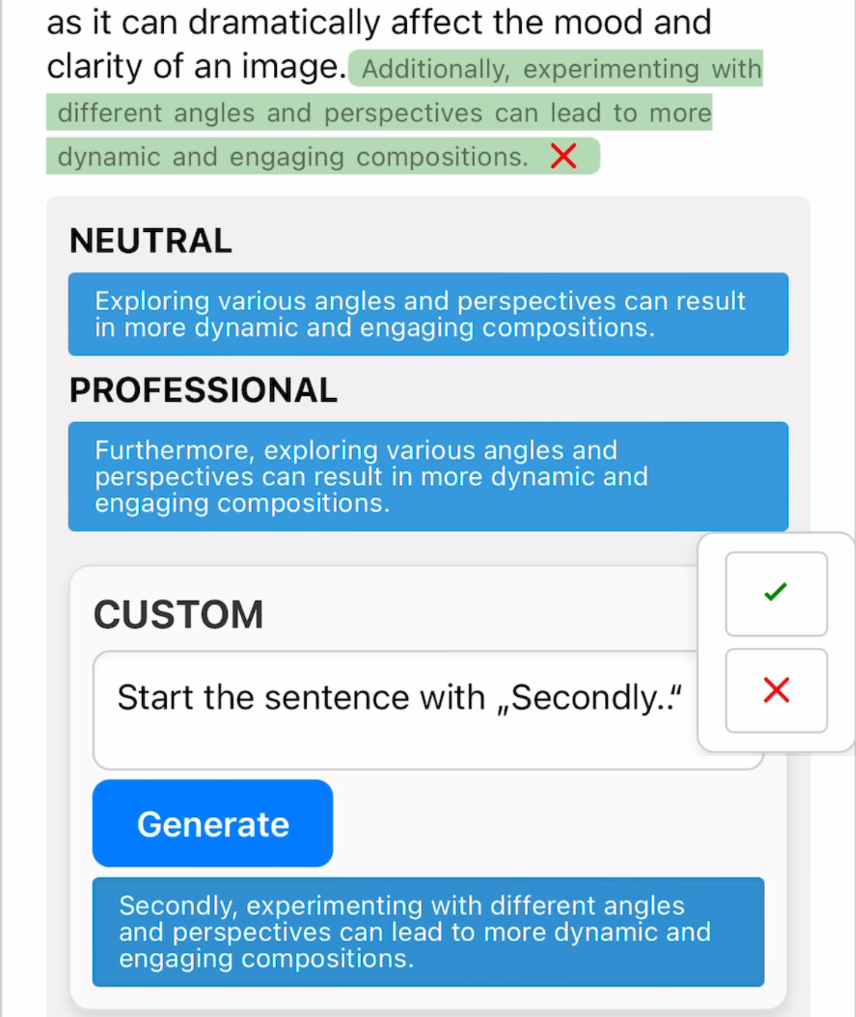}
        \caption{Long-pressing a full sentence presents users with a neutral and professional tone adjustment, along with the option to enter a custom prompt for further refinement.}
        \label{fig:long_press_sentence}
     \end{subfigure}
     \caption{Our long-press feature allows users to request synonyms on a word level (a) or tone adjustments on a sentence level (b). When words or sentences are replaced by selecting an alternative, they swap places to allow users to revert their action.}
     \Description{This figure shows two examples of longpress features designed for text editing. Image (a): The left-hand side displays a text passage in a text editor. When a user long-presses on the word 'grappled,' a suggestion box appears underneath, showing four alternative synonyms: 'struggled,' 'wrestled,' 'battled,' and 'fought.' The user can select one of these suggestions to replace the word 'grappled.' There is a checkmark and an 'X' icon in the bottom-right corner of the screen, allowing the user to confirm or reject their changes. Image (b): The right-hand side shows a different part of the text editor, where the user has long-pressed a full sentence. The interface presents three options for tone adjustment: 'Neutral,' 'Professional,' and 'Custom.' Each option includes a tone-specific sentence replacement, such as 'Furthermore, exploring various angles and perspectives can result in more dynamic and engaging compositions' for the 'Professional' option. The 'Custom' option allows the user to input a new sentence manually. A 'Generate' button is available for applying the changes, and like in (a), checkmark and 'X' icons provide confirmation or rejection options.}
     \label{fig:long_press}
\end{figure*}

\subsubsection{Sentence-Snap}
Pre-study findings showed that users often prefer generating one complete sentence at a time. 
To support this, we introduced the ``Sentence-Snap'' feature. 
When users perform a rapid spread gesture and quickly lift their fingers, the generation process halts after one complete sentence. 

\subsubsection{Longpress}
\label{sec:longpress}
In response to pre-study feedback, we introduced the ``Longpress'' feature, a pop-up that enables users to modify generated words and sentences (\cref{fig:long_press}). 
This is triggered \revision{by} touching a bubble for 500\,ms. %
This design also illustrates how the gestures can be combined with \revision{other UI} elements.

\paragraph{Adjusting Words (\cref{fig:long_press_words}):} When a long press is applied to a word bubble, the LLM is prompted (\cref{prompt:synonym}) to provide up to five synonyms, which are displayed directly in the text beneath the selected word as individual bubbles. 
Tapping on a synonym replaces the original word in the text. 
To dismiss all suggestions, the user can tap anywhere outside the synonyms.

\paragraph{Adjusting Sentences (\cref{fig:long_press_sentence}):} When a long press is applied to a sentence bubble, the LLM is prompted (\cref{prompt:rewrite_sentence}) to rewrite the sentence in two different tones: ``neutral'' and ``professional.'' \revision{We chose these} based on user feedback and inspired by industry applications. 
Rewritten sentences appear directly beneath the selected sentence in two separate boxes.
Tapping on a box replaces the sentence in the placeholder with the newly generated style, and a ``previous'' box appears, allowing the user to revert the change. 
\revision{Users can also} input their own specifications (``Custom'' textbox) \revision{that are then used as a prompt} (see \cref{prompt:custom_sentence}).

\section{Method}\label{sec:method}
\begin{figure*}
    \centering
    \includegraphics[width=\linewidth]{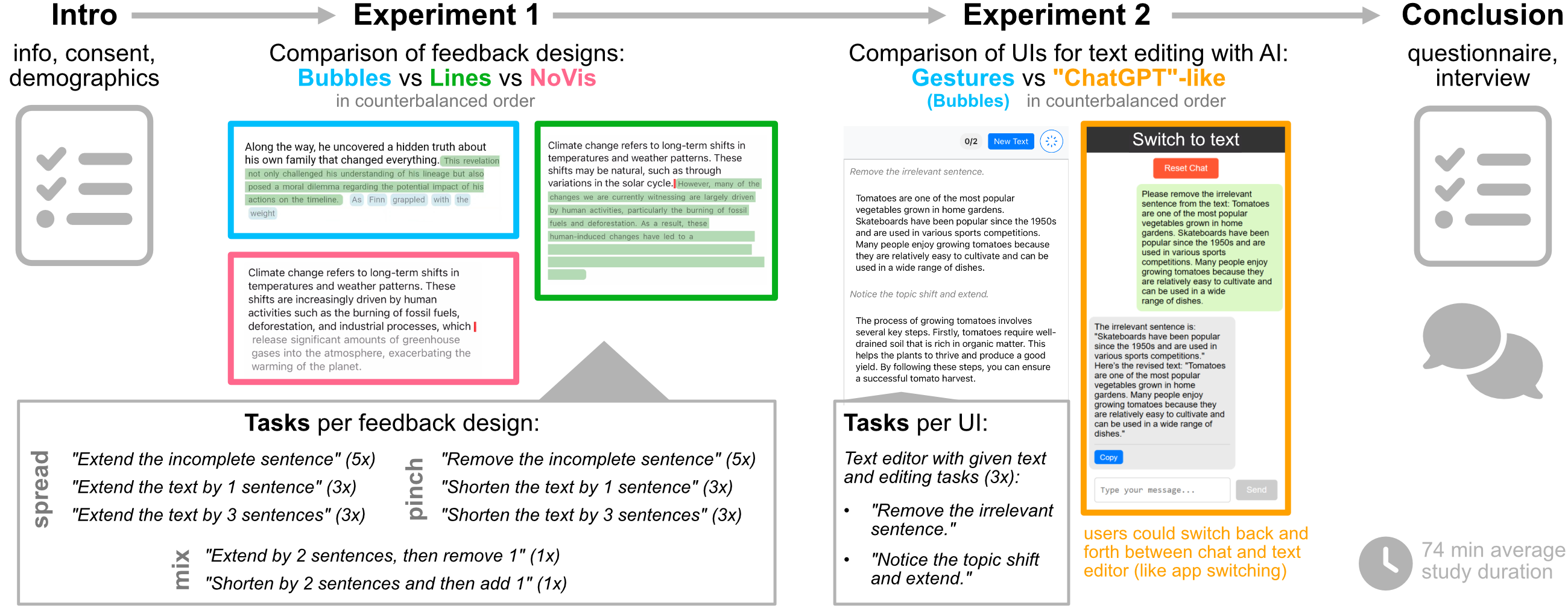}
    \caption{Overview of our user study design and procedure.}
    \Description{This figure provides an overview of the user study design and procedure, highlighting key steps, experiments, and tasks performed during the study, as they will be described in the following sections. The figure is structured into four main sections: Intro, Experiment 1, Experiment 2, and Conclusion. For each, it illustrates the content using screenshots and icons.}
    \label{fig:method_overview}
\end{figure*}

We conducted a within-subject \revision{lab} study to investigate touch gesture interaction for LLM-powered text length modification. \revision{Concretely, we compared} different feedback designs and evaluated \revision{the gestures} against a baseline chatbot interface. \cref{fig:method_overview} shows the study design and procedure.

\subsection{Experimental Design}
\revision{The study consisted of} two experiments and a ``hands on'' semi-structured interview.

\subsubsection{Experiment 1: Comparing Visual Feedback Loop Designs}
The goals for Experiment 1 were to observe how participants perceive the general concept, collect data on \spread{} and \pinch{} gesture execution, and compare three feedback designs.

\paragraph{Conditions}
The \textit{independent variable} was ``Feedback'' with three conditions: 
\visbubble{} -- the design shown in \cref{fig:teaser} and described in the previous section; 
\visline{} -- an alternative inspired by typical text selection markup, i.e. colouring the background of the text line (incl. without text when used in the placeholder role, cf.~\cref{sec:impl_text_length_ind}); 
and \visnone{} -- a baseline with no visual feedback beyond the text itself.
We selected these designs based on their ability to provide varying levels of feedback on text generation and modification (none vs length with \visline{} vs length + word count with \visbubble).
Their order was counterbalanced.

\paragraph{Measures}
As \textit{dependent variables}, we logged interaction metrics and collected subjective feedback through think-aloud protocols and questionnaires.

\paragraph{Tasks}
Both text extension and shortening were tested using short, neutral and unopinionated texts provided to participants as starting points (see \cref{sec:appendix_texts_exp1}).
We used the same texts for all conditions as the focus was not on reading and text comprehension, but repetitive execution of the gestures.

To explore different text length modifications, we asked participants to complete/remove one incomplete sentence (five repetitions each), extend/shorten the text by one full sentence (three repetitions each), extend/shorten the text by three sentences (three repetitions each), and perform a combination of extension and shortening tasks (twice).
The number of repetitions was chosen to balance task complexity with participant fatigue and potential learning effects.
To avoid additional confounds, we excluded the long-press feature (\cref{sec:longpress}) and disabled the phone's integrated keyboard.

\subsubsection{Experiment 2: Comparing Direct Touch Interaction Against a Chatbot UI}
This experiment aimed to provide deeper insights into how participants perceive the concept of controlling AI with touch gestures for text generation, compared to a typical chatbot UI. %

\paragraph{Conditions}
Interaction technique was the \textit{independent variable}, with two conditions: 
Direct Touch Interaction -- our proposed gesture-based text length modification using the Bubbles visualisation;
Chatbot Interaction -- a baseline imitating traditional chatbot-like interaction (\cref{fig:gpt_interface}).
\revision{We selected this baseline as a widely adopted, state-of-the-art approach for interacting with LLMs today, especially on mobile devices (e.g. see apps for ChatGPT\footnote{\url{https://openai.com/chatgpt/download/}} and Copilot\footnote{\url{https://play.google.com/store/apps/details?id=com.microsoft.copilot}})}. 
\revision{Suggestions and autocorrection were included via the normal phone keyboard.}
We counterbalanced the two conditions.

\paragraph{Measures}
As \textit{dependent variables}, we logged quantitative data (e.g. interaction logs) and qualitative data (e.g., observations, feedback, questionnaires).

\paragraph{Tasks}
We prepared short, unopinionated texts for participants to work with (\cref{sec:appendix_texts_exp2}).
They \revision{included} sentences clearly out of place (e.g. a topic shift from gardening to skating) or had missing parts (e.g. starting an enumeration but listing only one item).

Instead of giving quantifiable instructions on text length as in Experiment 1 (i.e. ``Remove \textit{one} sentence.''), this approach allowed us to give more abstract instructions. These required participants to read and understand the text (i.e. ``Remove the \textit{irrelevant} sentence.'').
Here we used different texts for the two conditions.
As in Experiment 1, tasks were considered completed when participants accepted the text length modification.

To explore more natural workflows in this more open task, we enabled the built-in mobile keyboard, and participants had access to the long-press feature.

\subsection{Apparatus and Materials}

\begin{figure*}[t]
     \centering
     \includegraphics[height=5.5cm]{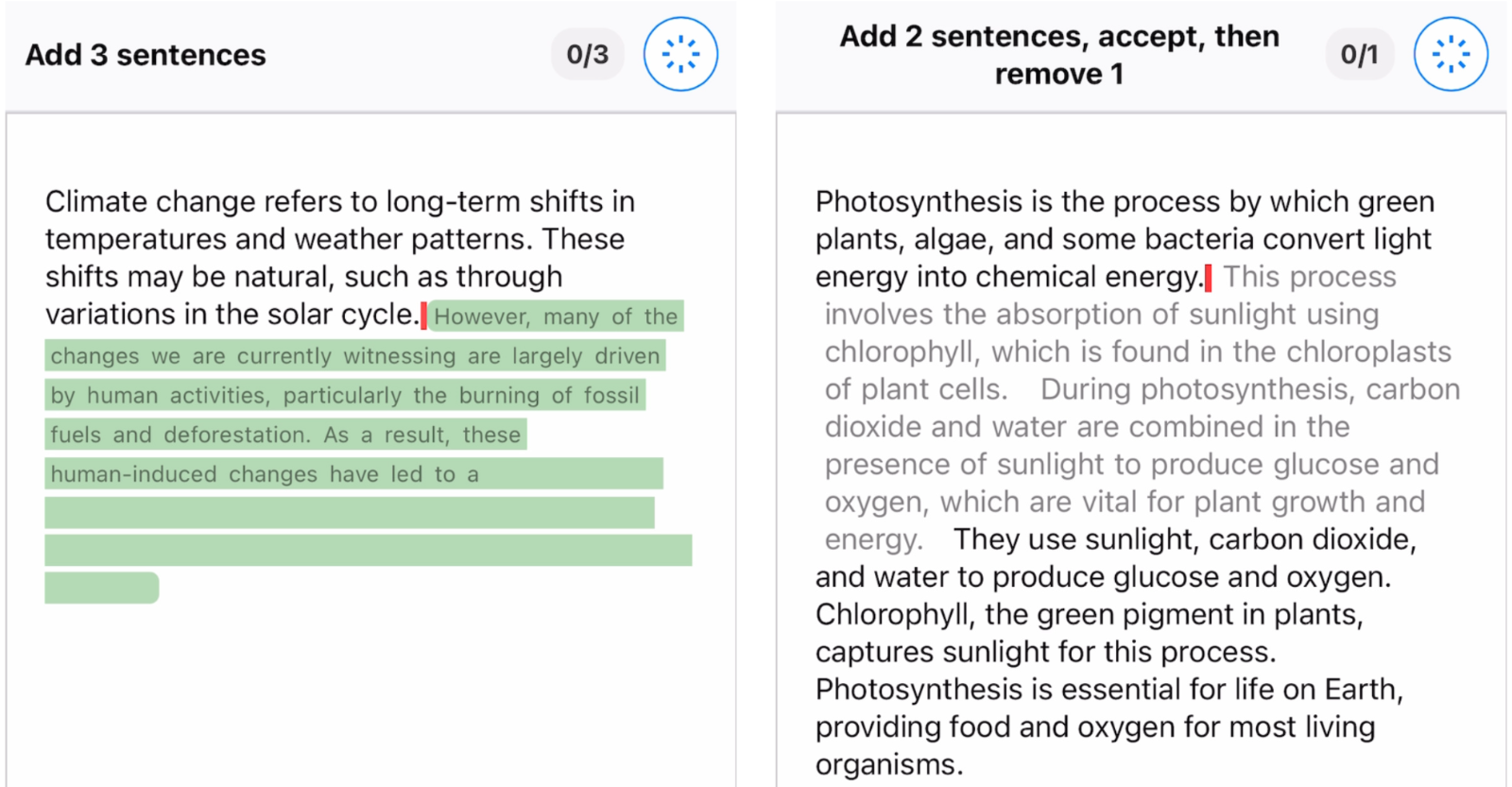}
     \caption{Examples of the UI in Experiment 1: In the \visline{} condition (left) coloured lines provide visual feedback on the change of text length. The design of the \visnone{} condition (right) offered no visual feedback beyond the text itself.}
     \Description{This figure shows two panels showing examples of the 'Lines' and 'NoVis' experimental conditions during Experiment 1 of the user study. Each panel demonstrates a moment of interaction where participants are tasked with editing text using different feedback methods. Panel (a): Lines Condition –- This panel shows the 'Lines' condition, where a coloured line is used to visualise changes in text length. The participant is tasked with adding three sentences to a text about climate change. The red cursor indicates the current insertion point. Green lines appear under newly generated text, highlighting the length of the text extension. The visual feedback helps the user keep track of the amount of text being added in real-time. Panel (b): NoVis Condition –- This panel illustrates the 'NoVis' condition, where no visual feedback is provided other than the text itself. The task asks the participant to add two sentences, accept the changes, and then remove one sentence. The red cursor shows where the participant is editing the text on photosynthesis. Without additional visual cues, participants must rely solely on the text to track their changes.}
     \label{fig:vis_conditions}
\end{figure*}

\subsubsection{Comparative Designs} %
\paragraph{Experiment 1}
For Experiment 1, we implemented two additional designs:
\visline{} \revision{indicated text-length as} a continuous green bar for generation, and a red bar \revision{for removing text}, but it had no placeholders for \revision{individual} words (\cref{fig:vis_conditions} left).
NoVis showed no text-length indicators at all (\cref{fig:vis_conditions} right).
In all three visualisations, generated text appeared as soon as the token stream arrived.

\begin{figure}
    \centering
    \includegraphics[height=7cm]{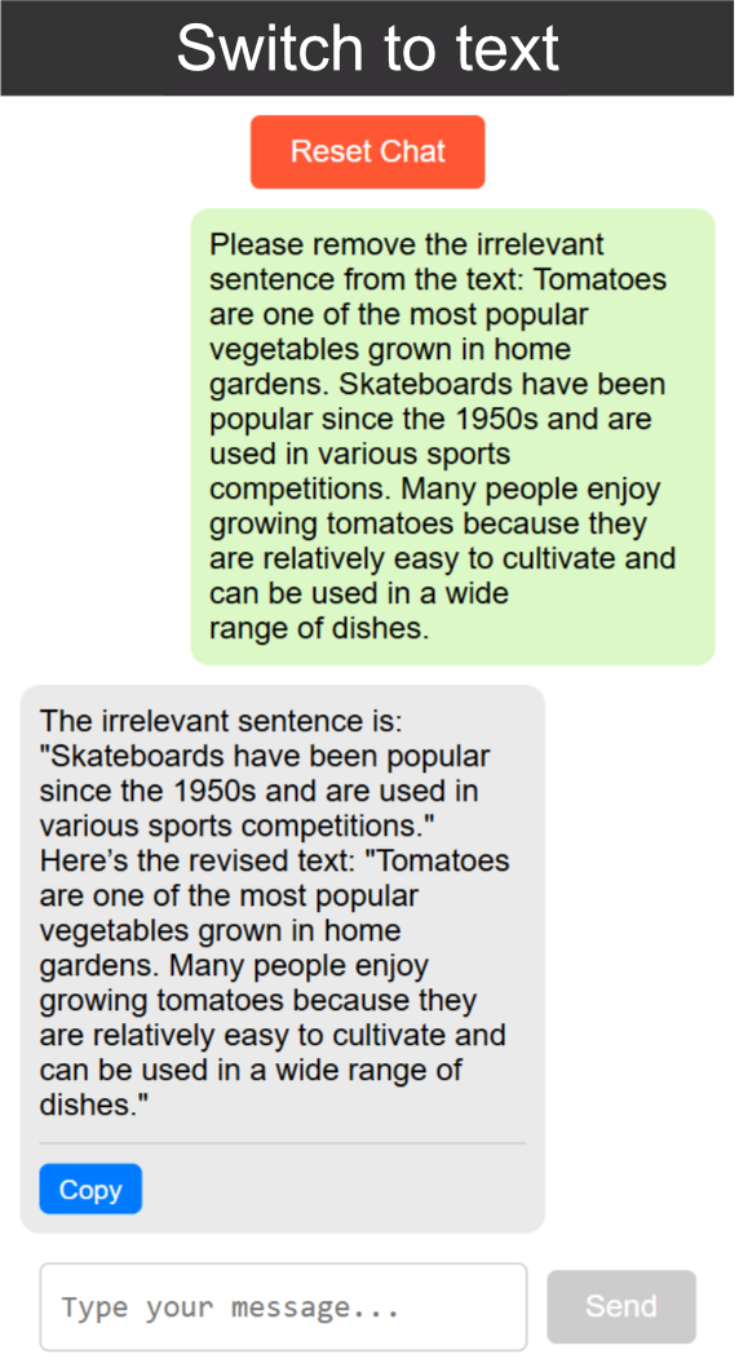}
    \caption{Our imitation of a chatbot-like interface, resembling the ChatGPT app. The green text bubble contains the user's prompt. The chatbot returns a grey text bubble, where the irrelevant sentence has been removed. Users can copy the revised text using the blue 'Copy' button. A chat input field and 'Send' button allow further interaction.}
    \Description{This figure shows a mobile interface designed to imitate a chatbot-like interaction, resembling the ChatGPT app. The interface includes two sections: Upper section: A green text bubble provides a task prompt, instructing the chatbot to 'remove the irrelevant sentence from the text.' The text to be edited follows, discussing tomatoes and skateboards. The instruction asks the chatbot to remove the sentence about skateboards, which is not relevant to the topic about growing tomatoes. Lower section: After processing the request, the chatbot returns a grey text bubble with the revised text. The irrelevant sentence about skateboards has been removed, and the corrected sentence about tomatoes is provided. A blue 'Copy' button is displayed below the revised text, allowing the user to copy the corrected version. At the bottom, there is a standard chat input box, where the user can type a message, along with a 'Send' button to interact with the chatbot.}
    \label{fig:gpt_interface}
\end{figure}

\paragraph{Experiment 2}
This experiment used our prototype with \visbubble.
For comparison, the same task was also completed using a chatbot-like UI, similar to the ChatGPT app.
To ensure consistent data logging across conditions and avoid differences in latency and logging frequency, we implemented the chatbot UI directly in our web app (\cref{fig:gpt_interface}), using the same LLM as for the gestures (\cref{sec:backend}). 
Participants could easily switch between the text editor and chatbot via a button at the top, making this ``simulated'' context switch faster than actual switching between separate apps.

\subsubsection{Study Environment}
\begin{figure}
    \centering
    \includegraphics[height=5cm]{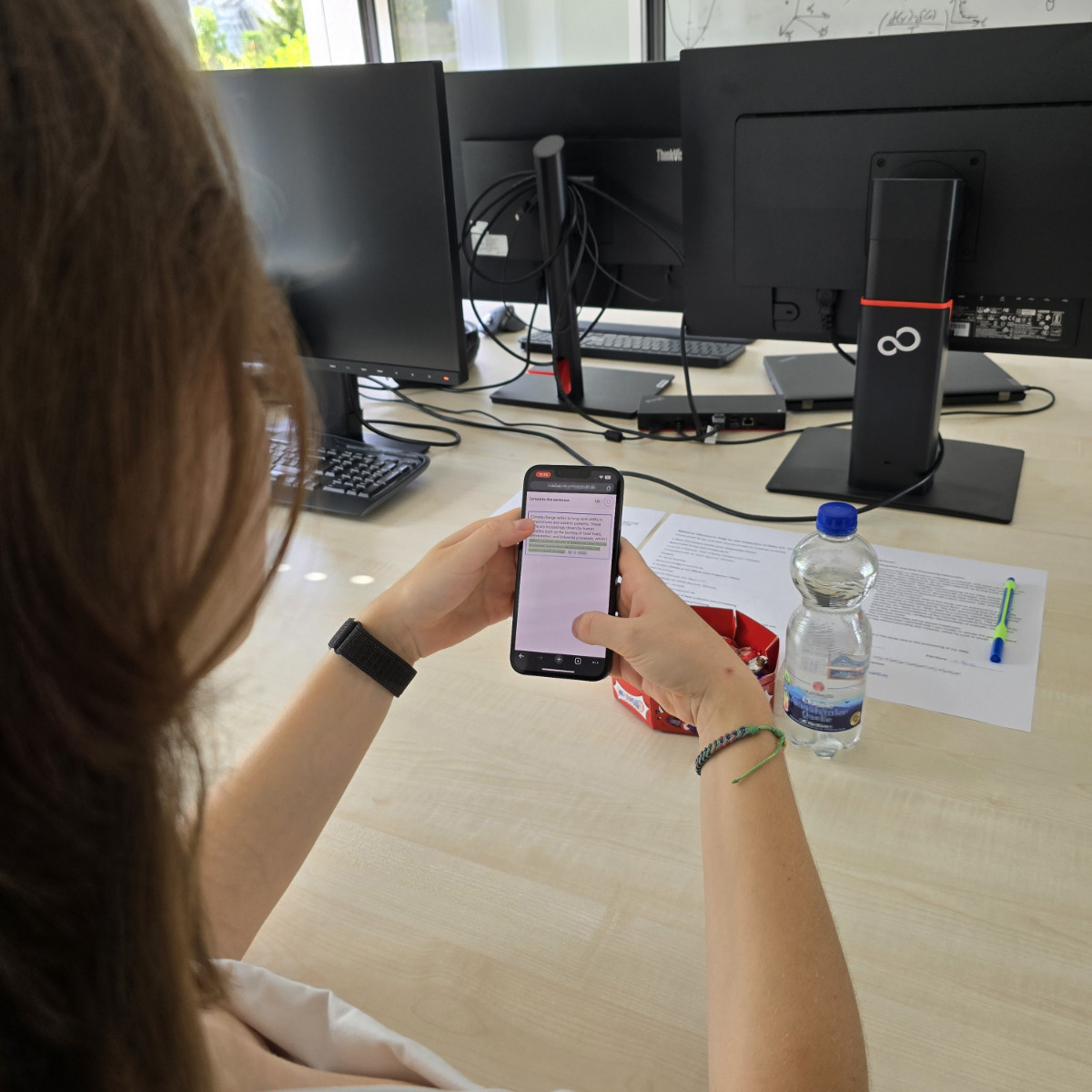}
    \caption{A participant performing the \spread{} gesture using our prototype web app. The participant provided consent to publish this figure.}
    \Description{This figure shows a participant interacting with a mobile device, performing the '\spread{}' gesture using the prototype web app. The participant is seated at a desk in an office environment with computer monitors and other office supplies in the background. The mobile device is being held with both hands, and the participant's fingers are spread apart on the screen, indicating the gesture used to control text generation. A bottle of water and a sheet of paper with study-related materials are visible on the desk.}
    \label{fig:lab_study}
\end{figure}

We conducted the study in person at our university lab in a neutral office environment with minimal distractions. 
Participants were seated comfortably, with both arms resting on a table while holding the device (\cref{fig:lab_study}).

We used an iPhone 14 with a screen resolution of 2532 × 1170 pixels and a screen size of 6.06\,in (153.924\,mm) running iOS 17.6.1. 
At the start of each experiment, the experimenter opened our web app using the Chrome Browser, and started the screen recording.

\subsubsection{Experimental Software and Web Application} %
\begin{figure}
    \centering
    \includegraphics[height=7cm]{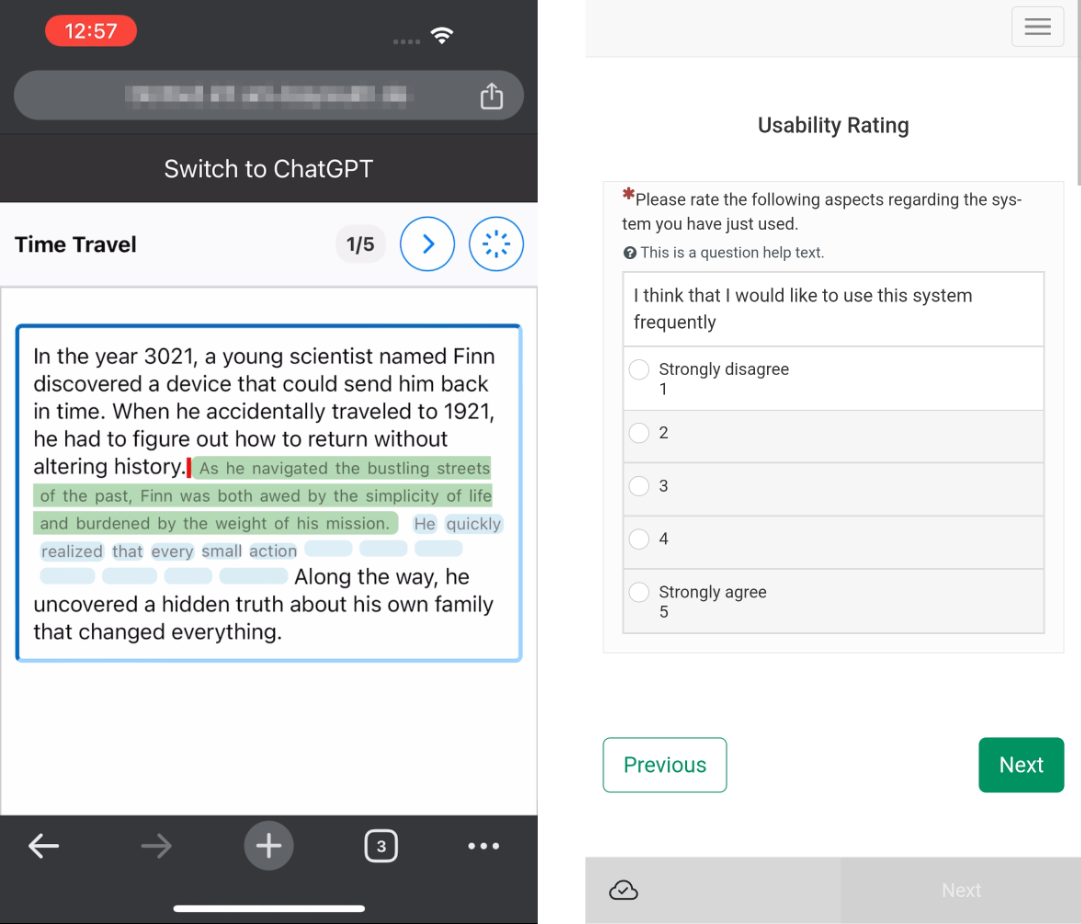}
    \caption{Two study elements showing the screen participants saw during the study with our web app opened inside a browser (URL blurred for anonymity), and a screenshot of one part of our questionnaire with UI elements for study progression. A digital clock in red indicates that screen recording was active.}
    \Description{This figure displays two screenshots representing key elements of the user study interface. Left panel: The left-hand image shows the screen participants viewed during the study while interacting with the web app inside a browser (with the URL blurred for anonymity). The app is presented in a Google Chrome mobile browser and displays a text-editing task from the prototype. In this specific instance, the participant is engaged with a semi-structured interview where they modify a passage about time travel by interacting with highlighted text. The red-highlighted digital clock at the top indicates that screen recording is active. Other UI elements, including task progression indicators and buttons for skipping to the next task or resetting the prototype, are also visible. Right panel: The right-hand image shows a section of the usability questionnaire that participants filled out as part of the study. The UI includes a rating scale from 1 (strongly disagree) to 5 (strongly agree) for the statement 'I think that I would like to use this system frequently.' Navigation buttons ('Previous' and 'Next') allow participants to move between sections of the questionnaire.}
    \label{fig:study_elements}
\end{figure}

We hosted our prototype (\cref{sec:implementation}) locally on our server. 
It was embedded into a framework that handled the study logic, including briefings and in-app questionnaires, and managed counterbalancing and study progression (\cref{fig:study_elements}). %

\begin{table*}
\footnotesize
\begin{tabular}{cccccccc}
\toprule
\textbf{P} & \textbf{Age} & \textbf{Gender} & \textbf{Occupation} & \textbf{WPM} & \makecell{\textbf{Experience with} \\ \textbf{Generated Texts}} & \textbf{Frequency of AI Usage} & \makecell{\textbf{Familiarity with LLMs} \\ \textbf{(Likert scale, 1/low to 5/high)}}  \\
\midrule
P1 & 22 & Male & Student & 30 & Yes (ChatGPT) & Occasionally & 5 \\
P2 & 20 & Male & Apprentice & 36 & Yes (ChatGPT) & Rarely & 4 \\
P3 & 20 & Male & Student & 38 & No & Rarely & 2 \\
P4 & 20 & Male & Student & 26 & No & Rarely & 1 \\
P5 & 22 & Male & Student & 53 & Yes (study related) & Occasionally & 5 \\
P6 & 20 & Male & Student & 37 & No & Rarely & 2 \\
P7 & 24 & Male & Student & 44 & Yes (ChatGPT) & Occasionally & 5 \\
P8 & 20 & Female & Student & 29 & Yes (ChatGPT) & Occasionally & 4 \\
P9 & 21 & Female & Student & 35 & No & Rarely & 1 \\
P10 & 48 & Female & Technical Assistant & 22 & No & Rarely & 1 \\
P11 & 60 & Female & Technical Assistant & 16 & No & Rarely & 1 \\
P12 & 19 & Male & HS Graduate & 34 & Yes (ChatGPT) & \revision{Rarely} & 2 \\
P13 & 59 & Female & Midwife & 10 & No & Rarely & 1 \\
P14 & 21 & Male & Student & 50 & Yes (ChatGPT) & Occasionally & 5 \\
\bottomrule
\end{tabular}
\caption{Overview of the participants.}
\Description{This table gives an overview of the participants.}
\label{tab:participants}
\end{table*}

\subsection{Participants}
We recruited 14 participants (5 female, 9 male) through our university network. 
Criteria for participation were high English proficiency and the ability to perform touch gestures on mobile devices. 
To ensure a more diverse range of perspectives and assess the accessibility of our concept across different demographics, we selected participants from distinct age groups and occupational backgrounds.
All participants were right-handed. %
While writing on mobile devices, most participants reported to use both hands and thumbs (11), two write one-handed with the thumb, and one with their index finger. 
All use their mobile devices (11 iOS, 3 Android) at least several times a day, mainly for writing messages (13), notes, comments and TODOs (3), and browsing the internet (3).
\cref{tab:participants} provides an overview. %

\subsection{Procedure}

Participants \revision{received} study documents in accordance with our university's regulations. %
After giving informed consent, participants filled out a demographic questionnaire.

We explained the experiment and demonstrated interaction with our prototype. 
This covered using gestures to adjust text length and the experimental conditions.
We encouraged participants to interact with the device in whichever way felt most comfortable. %
Furthermore, we asked them to ``think aloud'' while interacting. %
The phone was cleaned with disinfectant wipes between participants.

To \revision{assess} manual typing speed and help participants familiarize themselves with the device, they first completed a one-minute typing test, typing random English words. 
We answered all questions that came up.

The study took around 74 minutes on average.
Participants were compensated at a rate of €\,5 per (started) 20 minutes. %

\subsubsection{Experiment 1} %
\label{sec:procedure_exp1}
The same three tasks (extend, shorten, combination) were performed on the same texts (see \cref{sec:appendix_texts_exp1}) for all feedback designs, structured as follows:

First, participants extended the text by completing one incomplete sentence (5 repetitions), adding one sentence (3 repetitions), and adding three sentences (3 repetitions).
Second, they removed one incomplete sentence, one full sentence, and three sentences, repeating each task as previously.
Finally, participants performed combinations: 
In the first, they added two sentences and then removed one; 
in the second, they removed two sentences and then added one. 
These combination subtasks were performed only once.

After the final subtask \revision{per condition}, participants answered the in-app questionnaires and put the phone down \revision{for a} short break.
They then repeated the tasks with the next counterbalanced condition. 
\revision{Finally,} Experiment 1 concluded with a questionnaire and an extended break. %

\subsubsection{Experiment 2} %
\label{sec:procedure_exp2}
Here we instructed participants to edit a provided text by either removing sentences that did not fit the surrounding context or extending the text to fill logical or semantic gaps.
Beyond these abstract instructions, we did not specify the location of the text modifications or the amount of text to be removed or added. 
Thus, participants had to determine for themselves when they had changed the text to their satisfaction.
\revision{This was repeated for} three different texts per condition (see \cref{sec:appendix_texts_exp2}).
After the first condition, we asked participants to answer in-app questionnaires and put the phone down \revision{for} a short break, \revision{before continuing} with the second condition.
\revision{At the end, participants answered} a questionnaire regarding their experience with both conditions.

\subsubsection{Semi-Structured Interview}
\revision{The study sessions concluded with} a semi-structured interview to gather detailed feedback. 
During the interview, participants were free to explore our prototype (with long-press and keyboard enabled) without any prescribed tasks.
They could also switch to the chatbot UI.
If they wished, we provided a selection of short creative story texts (see \cref{sec:short_stories}) as inspiration, though they were free to write anything they wanted.
This allowed them to directly demonstrate what they were referring to during the interview. %

\section{Results}
Here we report on the study results.
Across both experiments, we recorded 1706 gestures from the 14 participants.

\subsection{Time}
\label{ssec:time}
\begin{figure*}[t]
    \centering
    \includegraphics[width=1\linewidth]{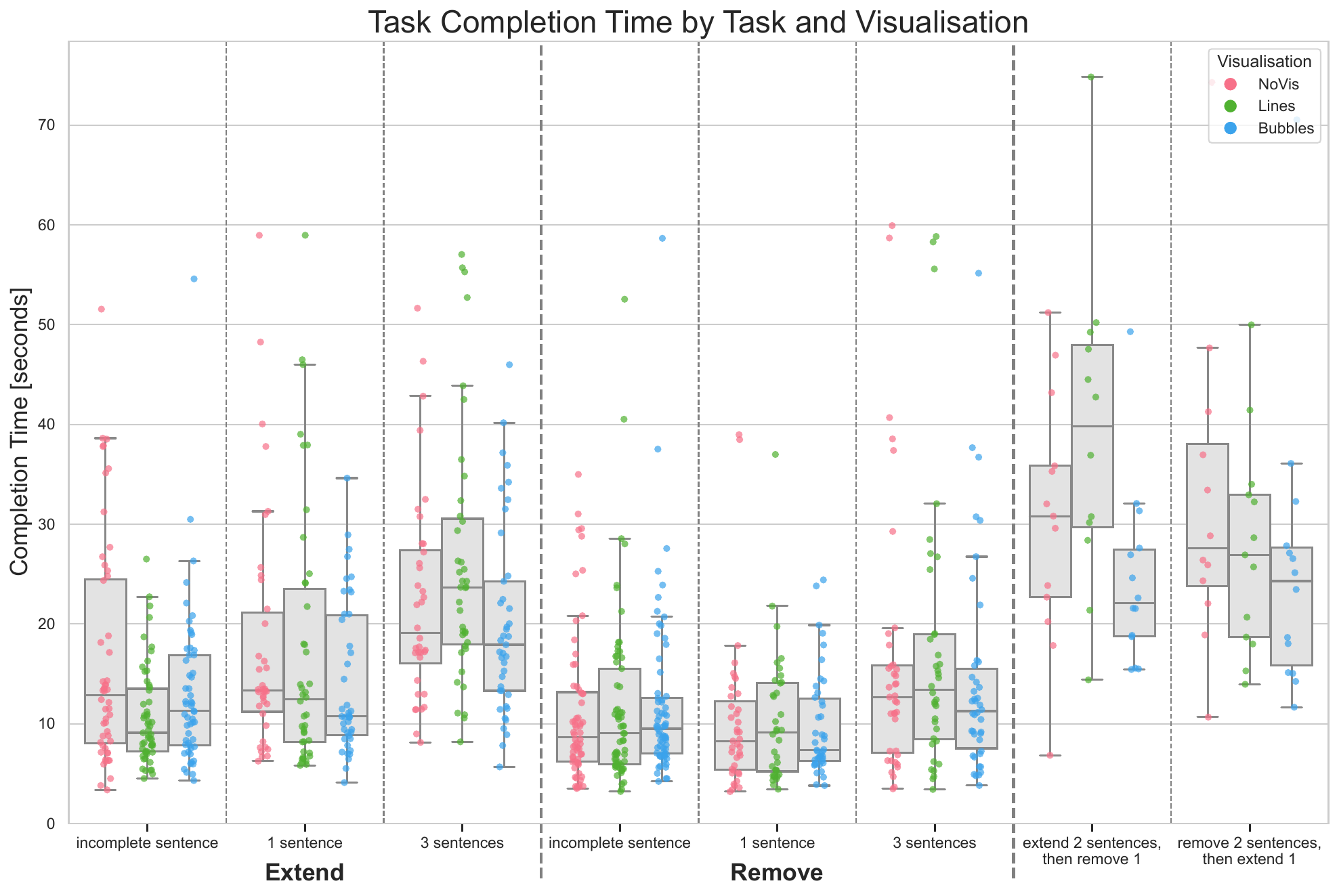}
    \caption{Task completion times for all subtasks in Experiment 1, segmented by visualisation condition (NoVis, Line, and Bubbles). Each subtask is represented along the x-axis, with the y-axis showing the completion time in seconds. Box plots display the completion times for each visualisation condition, with individual data points plotted as dots.}
    \Description{This figure shows a box plot illustrating task completion times across different tasks, subtasks, and visualisation conditions from Experiment 1. The x-axis represents the task and subtask combinations, while the y-axis shows completion time in seconds. Each subtask is labelled along the x-axis (e.g., extend - incomplete sentence, extend - 1 sentence, etc.), representing various tasks the participants completed during the study. The y-axis measures the time taken to complete each task, ranging from 0 to 70 seconds. Three colours represent the different visualisation conditions: Red represents NoVis, Green represents Lines visualisation, Blue represents Bubbles. Each box plot indicates the distribution of completion times for each visualisation across the different subtasks. Individual data points are plotted as dots to provide a clearer view of task variability.}
    \label{fig:exp1_times}
\end{figure*}

In Experiment 1 we asked participants to repeatedly perform spread and pinch gestures to generate and remove text.
The mean time to complete this task (i.e. performing all gestures), was 16.38\,s when no visualisation was used, \secs{16.30} for Lines and \secs{14.41} for Bubbles.
These differences were significant as follows (\cref{tab:sig_tests}, row 1): 
\revision{With Bubbles,} participants completed the tasks significantly faster than with Lines (-\secs{1.76}) or without a visualisation (-\secs{2.62}).
\cref{fig:exp1_times} shows each subtask as a box blot.

\revision{In Experiment 2}, with a mean completion time of \secs{56.35}, Bubbles was twice as fast as the chatbot UI, which took people \secs{134.86}.
This difference is significant (\cref{tab:sig_tests}, row 2).

The mean execution time of one extend gesture was \secs{5.23} for NoVis, \secs{4.59} for Lines, and \secs{3.88} for Bubbles.
Gestures to remove text took an average of \secs{2.73} for NoVis, \secs{2.59} for Lines, and \secs{2.52} for Bubbles.
\revision{See \cref{tab:time} for more details.}

\begin{table}[t]
\footnotesize
\centering
\begin{tabular}{llccc}
\toprule
\textbf{Task} & \textbf{Variable} & \textbf{Time} & \textbf{SD} & \textbf{Median} \\
\midrule
\multicolumn{5}{c}{\textbf{Experiment 1 - Visualisation}} \medskip\\
Overall & No Vis & 16.38 & 11.80 & 12.95 \\
 & Lines & 16.30 & 12.36 & 12.16 \\
 & Bubbles & 14.41 & 8.96 & 11.53 \medskip\\
Extend & No Vis & 18.48 & 11.53 & 13.89 \\
 & Lines & 17.16 & 12.22 & 13.24 \\
 & Bubbles & 15.58 & 8.91 & 12.87 \medskip\\
Shorten & No Vis & 12.27 & 9.87 & 9.52 \\
 & Lines & 12.79 & 10.36 & 9.78 \\
 & Bubbles & 12.02 & 8.54 & 9.24 \medskip\\
Combination & No Vis & 31.48 & 14.26 & 29.60 \\
 & Lines & 33.18 & 14.44 & 30.76 \\
 & Bubbles & 25.12 & 12.04 & 23.03 \medskip\\
\midrule
\multicolumn{5}{c}{\textbf{Experiment 2 - Interaction}} \medskip\\
Combined Task & Gesture & 56.35 & 30.87 & 51.03 \\
 & CUI & 134.86 & 58.05 & 125.94 \medskip\\
\midrule
\multicolumn{5}{c}{\textbf{Gesture Execution Time}} \medskip\\
Extend & No Vis & 5.23 & 3.13 & 4.71 \\
 & Lines & 4.59 & 2.45 & 4.33 \\
 & Bubbles & 3.88 & 2.20 & 3.61 \medskip\\
Shorten & No Vis & 2.73 & 1.45 & 2.22 \\
 & Lines & 2.59 & 1.51 & 2.00 \\
 & Bubbles & 2.52 & 0.87 & 2.41 \\
\bottomrule
\end{tabular}
\caption{Times measured in the study.}
\Description{This table gives an overview of the task times during the user study.}
\label{tab:time}
\end{table}

\subsection{\revision{Use of Space}}
\begin{figure}[t]
    \centering
    \includegraphics[height=7cm]{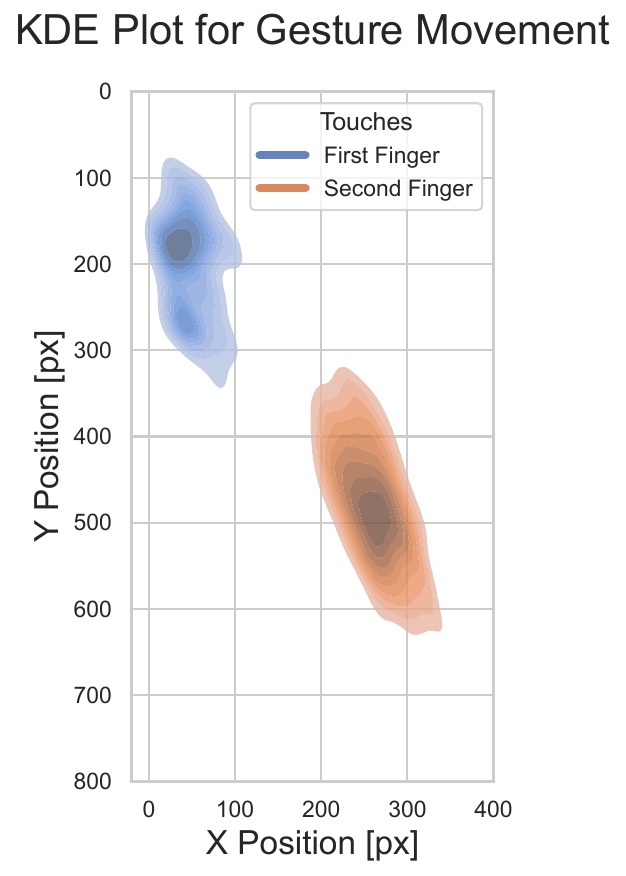}
    \caption{Density plot of all finger positions that were logged during the user study.}
    \Description{This figure is a Kernel Density Estimate (KDE) plot, visualising the density of all finger positions during gesture movements in the user study. The plot contains two density regions, each representing the positions of the first and second fingers during gesture input. The blue contour represents the distribution of the first finger’s touch positions, which are concentrated in the upper-left portion of the plot, around the coordinates (100, 150) in X and Y pixels, respectively. The orange contour indicates the density of the second finger’s positions, which are located in the lower-right part of the plot, around the coordinates (250, 500).The X-axis represents the horizontal position of the touches in pixels, and the Y-axis represents the vertical position. Darker areas within each density region indicate where finger positions are more concentrated during the gestures.}
    \label{fig:finger_kde}
\end{figure}

The average finger distance when generating one sentence with the spread gesture was 229 pixel or 38.1\,mm. %
The average with the pinch gesture was 130 pixel or 21.6\,mm. %
\cref{fig:finger_kde} shows \revision{the finger locations.}
Participants used 1.88 (SD 1.77, median 1) subgestures (defined as taking the fingers off the screen to readjust) to complete one sentence, 1.73 (SD 1.35, median 1.0) to add a full sentence, and 3.89 (SD 2.70, median 3) to add three.
Similarly, they used 2.30 (SD 2.55, median 2) subgestures to remove one incomplete sentence, 1.38 (SD 0.64, median 1) to remove one sentence, and 2.92 (SD 1.68, median 2) to remove three.

\begin{figure}[t]
    \centering
    \includegraphics[width=1\linewidth]{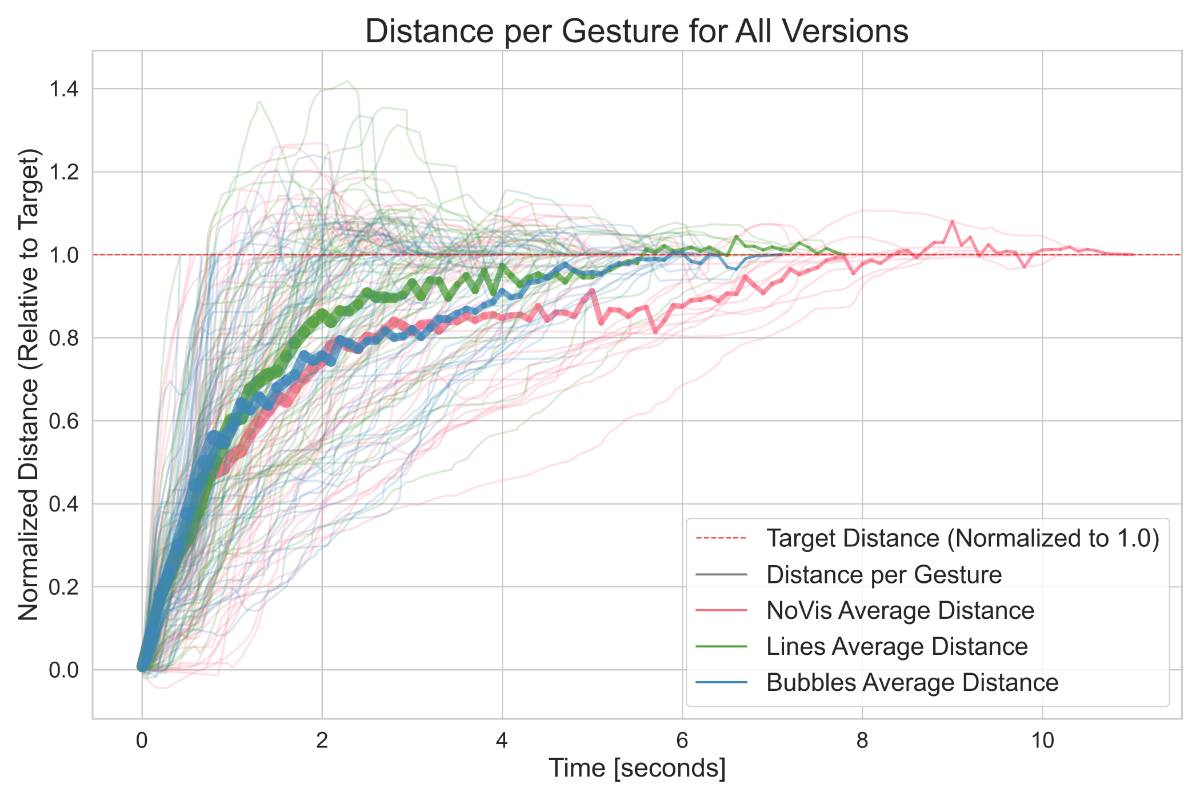}
    \caption{The progression of normalized distance (relative to the target) over time for the visualisations in Experiment 1. The dashed red line represents the target distance, which is defined as the distance the fingers needed to be moved apart to create the intended amount of text. The pink, green, and blue lines represent the average distance at each point in time for \visnone{}, \visline{}, and \visbubble, respectively. The thickness of the lines indicates the number of samples included. Individual traces are displayed in light background lines.}
    \Description{This line graph illustrates how the distance between fingers, normalized relative to a target value, changes over time under different visual conditions tested in Experiment 1. The x-axis represents time in arbitrary units, while the y-axis represents normalized distance (relative to a target value of 1.0). A dashed red line indicates the target distance, which corresponds to the ideal finger movement required to achieve the desired text change. Three average distance lines are drawn: a pink line for the NoVis condition, a green line for the Lines condition, and a blue line for the Bubbles condition. The thickness of the lines indicate the number of samples included, which decreases gradually. Numerous faint individual traces are shown in the background, representing the variations in gesture distances for different trials under each condition. The lines for Bubbles reaches the dashed red line the fastest. NoVis reaches this target line the latest and overshoots.}
    \label{fig:normdist_time_all_avg}
\end{figure}

\subsection{Anatomy of Touch Gesture Text Generation}

\revision{To initiate the gestures, participants placed one finger at either the end or middle of a sentence, with the second finger positioned closer for the \spread{} gesture (280 px on average, SD 77.5, median 274) than for the \pinch{} gesture (343 px on average, SD 70.5, median 351). Also see \cref{fig:first_touches} in \cref{sec:appendix_figs}.}

An analysis inspired by human pointing dynamics \cite{mouse_mueller2017} \revision{revealed} distinct phases during the \spread{} gesture (\cref{fig:normdist_time_all_avg}): %
Participants rapidly spread their fingers to generate text, \revision{with 80\% of the target distance (where the task was completed) covered in the first half of the movement time.}
Afterward, the movement slowed down as the text approached the intended length.

\begin{figure*}[t]
     \centering
     \begin{subfigure}[t]{0.49\textwidth}
        \centering
        \includegraphics[width=0.95\linewidth]{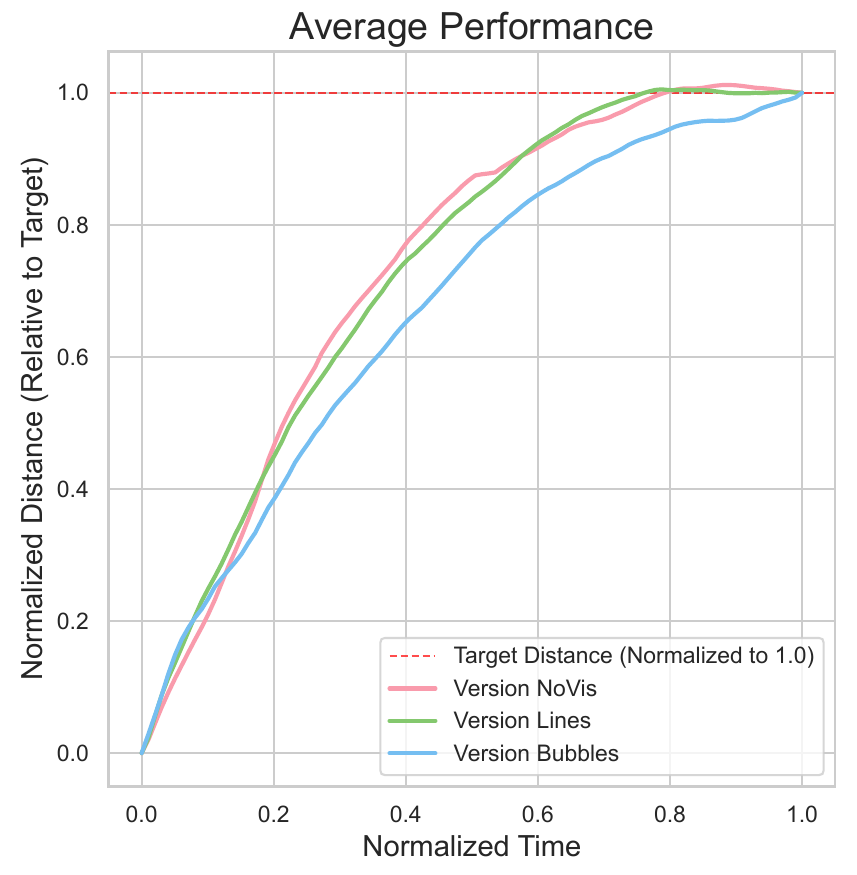}
        \caption{The complete average movement to generate the targeted amount of text with the \spread{} gesture as a function of distance (relative to the target text length) over normalised time.}
        \label{fig:overshoot1} 
     \end{subfigure}
     \hfill
     \begin{subfigure}[t]{0.49\textwidth}
        \centering
        \includegraphics[width=0.9555\linewidth]{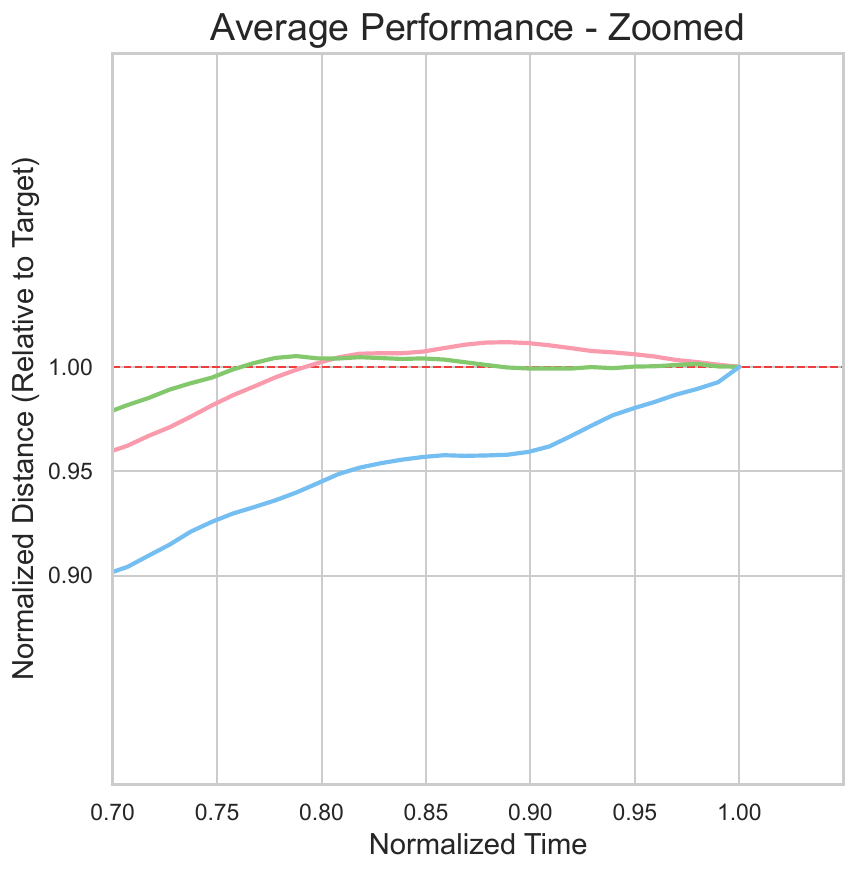}
        \caption{The movement from (a) zoomed in on the end of the movement. At around 80\% of the total movement time a visible overshoot occurs for both \visnone and \visline.}
        \label{fig:overshoot2}
     \end{subfigure}
     \caption{The average movement to generate the target amount of text with \spread{} gives insights into how this gesture was executed. The distance is normalised relative to the targeted amount of text (red dashed line) and the normalized time. Points above y=1 indicate an overshoot, where more text was being generated than intended. We obtain this intention directly from users, as they had to confirm the text length modification after performing the gesture.}
     \Description{This figure contains two line graphs comparing the performance of different visualisation conditions ('NoVis', 'Lines', and 'Bubbles') while using the \spread{} gesture to generate text. Left graph (a): This graph plots the complete average movement required to generate the targeted text length as a function of normalized distance (relative to the target text length) over normalized time. The x-axis represents normalized time (0.0 to 1.0), and the y-axis represents normalized distance (relative to the target distance of 1.0). The three visualisation conditions ('NoVis', 'Lines', and 'Bubbles') are compared, with each shown as a distinct line. The target distance is represented by a red dashed line, while the other lines show the performance of each condition. The graph demonstrates how each condition approaches the target distance over time. Right graph (b): This graph zooms in on the final portion of the movement, focusing on the last 20\% of the total movement time. The same three conditions ('NoVis', 'Lines', and 'Bubbles') are plotted, showing the movement beyond 0.70 normalized time. Around 80\% of the total movement time, a visible overshoot occurs in both 'NoVis' and 'Lines' conditions, where the lines exceed the target distance.}
     \label{fig:overshoot}
\end{figure*}

Without visual feedback, participants often overshot the intended text length, going beyond the distance needed to complete the task.
This overshooting was less prominent with the Lines visualisation and entirely absent with Bubbles, as shown in \cref{fig:overshoot}. 

For Bubbles, the target was approached with a more consistent velocity.
Note that in \cref{fig:overshoot}, both the distance to the target and the time have been normalized\revision{, making the} curve for Bubbles appear to show a delayed target reach compared to the others. 
However, in absolute terms, Bubbles \revision{achieved} the shortest execution time, as detailed in \cref{ssec:time} and \cref{fig:normdist_time_all_avg}.

Finally, regarding hand postures, many participants intuitively used both thumbs for the gestures, though some adjusted their technique depending on the text's position, switching between thumbs and index fingers. 

\subsection{Perception of Visualisation Techniques (Experiment 1)}
\label{sec:perception_exp1}
After each condition in Experiment 1, participants rated the system's usability with the SUS questionnaire, their perceived workload using the NASA-TLX, and nine additional 5-point Likert items.

These results show that Bubbles achieved the highest usability rating of 85.54, indicating ``excellent'' usability, whereas Lines achieved a ``good'' score of 76.96, and NoVis performed significantly (\cref{tab:sig_tests}, row 3) worse than both with an ``OK'' score of 63.46 \cite{susBangor2009}.
Similarly, participants perceived the lowest workload using Bubbles with a NASA-TLX Raw score of 1.976, followed by 2.154 for Lines, and 2.794 for NoVis, which was significantly higher than both visualisations (\cref{tab:sig_tests}, row 5).

\begin{figure*}
    \centering
    \includegraphics[width=.65\linewidth]{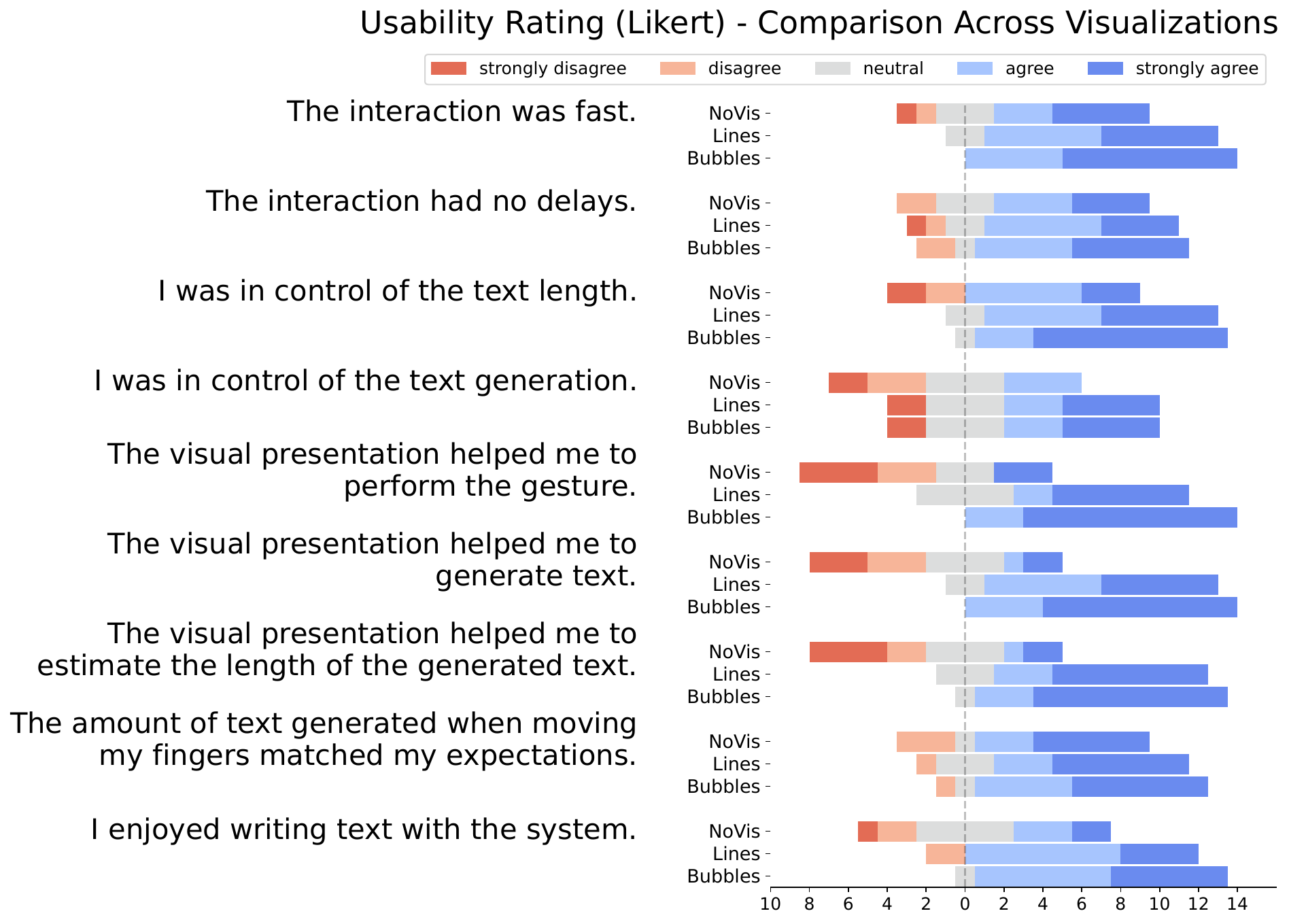}
    \caption{Likert results on participants' perception of interaction with our prototype and the visual presentations, rated after each condition in Experiment 1. These rating overall indicate that participants felt that their interaction with the system was fast and without delays, regardless of the visualisation technique, but comparatively, NoVis was clearly rated the lowest across all items.}
    \Description{This figure shows divergent bar charts for the Likert results on participants' perception of interaction with our prototype and the visual presentations, rated after each condition in Experiment 1. These rating overall indicate that participants felt that their interaction with the system was fast and without delays, regardless of the visualisation technique, but comparatively, NoVis was clearly rated the lowest across all items.}
    \label{fig:own_likert_exp1}
\end{figure*}

Further Likert items in Experiment 1 (see \cref{fig:own_likert_exp1}) overall indicate that participants perceived their interaction with the system as fast and without delays, regardless of the visualisation.
Comparatively, NoVis was clearly rated the lowest across all items.
While this is expected for questions regarding the support offered by visual representations, it is worth noting that both Lines and Bubbles also %
considerably increased participants' enjoyment of the system.
Comparing Bubbles and Lines, Bubbles was rated higher across all items, except for participants' perceived level of control over text length, where both visualisations were rated equally high. 

\subsection{Perception of Interaction (Experiment 2)}
\label{sec:perception_exp2}
In Experiment 2, participants rated the system's usability significantly higher (\cref{tab:sig_tests}, row 4) for the touch gestures (81, ``good'') compared to using the CUI designed to mimic ChatGPT (52.5, ``OK'').
Similarly, their perceived workload was significantly (\cref{tab:sig_tests}, row 6) lower when using touch gestures (2.060) compared to using ChatGPT (3.153). %

\begin{figure*}
    \centering
    \includegraphics[width=.65\linewidth]{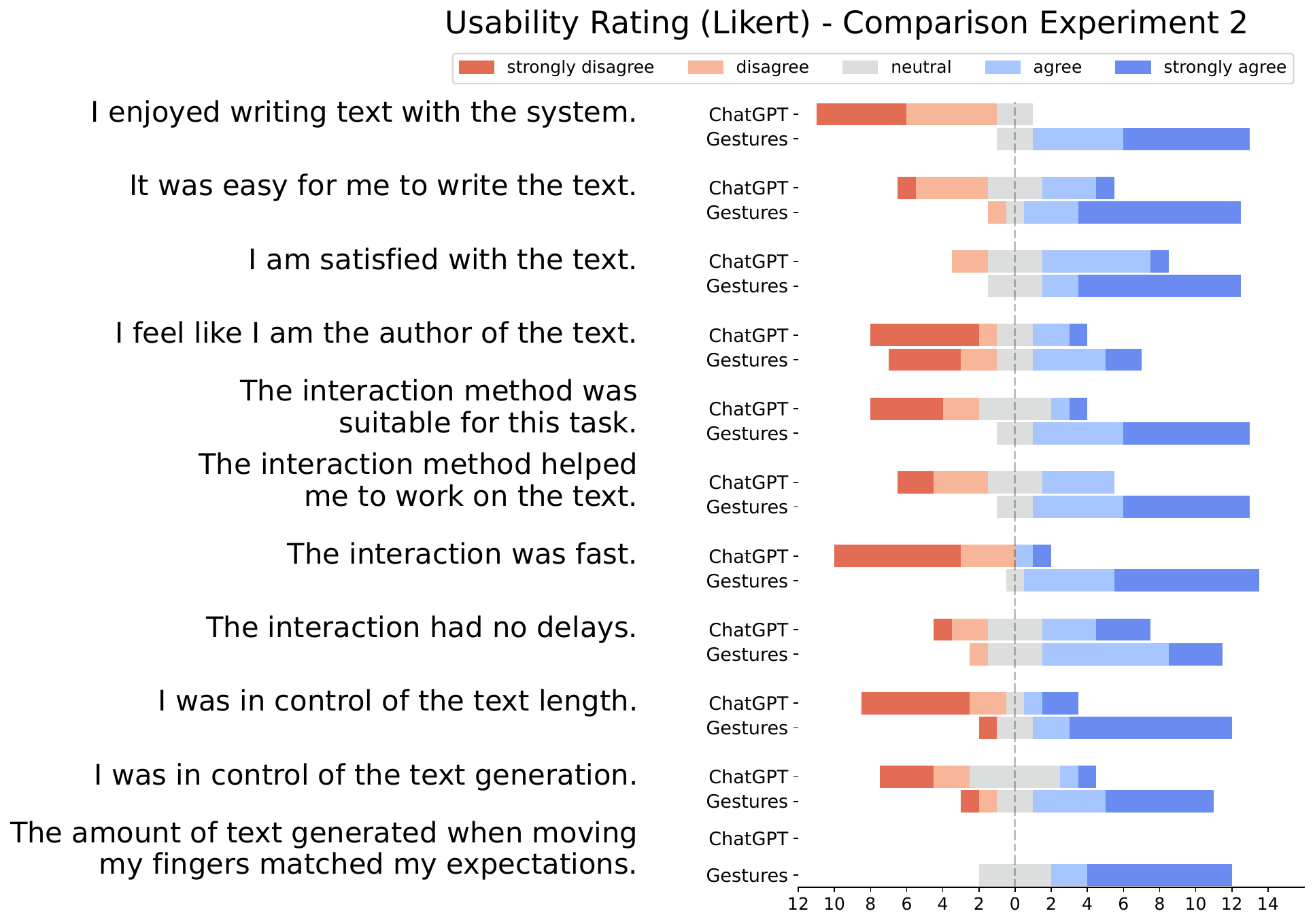}
    \caption{Likert results on participants' perception of interaction with our touch-based prototype compared to our CUI implementation, rated after each condition in Experiment 2. Most participants rated almost all items overwhelmingly positive when using our approach.}
    \Description{This figure shows divergent bar charts for the likert results on participants' perception of interaction with our touch-based prototype compared to our CUI implementation, rated after each condition in Experiment 2. Most participants rated almost all items overwhelmingly positive when using our approach. Only for the item that asked participants whether they felt like the author of the text, while still rating our mode higher, the responses were more mixed.}
    \label{fig:own_likert_exp2}
\end{figure*}

This trend is further reflected in the nine Likert-items for Experiment 2 (see \cref{fig:own_likert_exp2}), which participants rated after each condition.
Our gesture-based interaction approach scored higher than the ChatGPT-like CUI across all items.

\subsection{Subjective Feedback}
\label{sec:ss_interview_perception}
After both experiments, participants rated four Likert items on their overall experience with the gestures.

All but one ``neutral'' participant agreed (7) or strongly agreed (6), that ``the gesture controls felt intuitive.''
Similar, all but one ``neutral'' participant agreed (5) and strongly agreed (8), that ``the gesture controls felt natural''.
They ``would use this gesture-based text control of text generation feature in [their] daily tasks'' (4 ``neutral'', 8 ``agree'', 2 ``strongly agree'') and ``would recommend this gesture-based text control feature to others'' (1 ``neutral'', 3 ``agree'', 10 ``strongly agree'').
No one disagreed with any of these statements.

As the last part, we conducted semi-structured interviews. %
All but one participant %
picked Bubbles as their favourite, explicitly confirming the bigger picture observed in the preceding sections.

When asked about challenges, six participants reported no issues, one mentioned occlusion, and seven said they sometimes struggled to select the intended sentence. 
\revision{Difficulties include ``fat finger'' issues on the smaller screen (some were accustomed to larger devices), misunderstanding the interaction pattern (placing the cursor instead of tapping the sentence), and lack of individual adjustments for visual feedback delay and touch target offset. %
}
However, P8 added, ``you'd get into it over time''\revision{, which we also observed in other participants as the study progressed}.

\revision{Participants’ feedback on what they liked most about using gesture controls emphasized its intuitive, ``extremely smooth'' (P14) and natural feel.
They preferred the Bubbles visualisation for its clear separation of words, sentences, and deletions.}
In \cref{sec:perception_exp2}, we reported mixed feelings about authorship (\cref{fig:own_likert_exp2}). 
However, when reminded about the long-press feature in the interviews, 
\revision{11 out of 14 participants said this would enhance their sense of authorship.}

Multiple participants also noted that the visualisation effectively managed latency, with P7 remarking that they had ``hardly any feeling of latency''. %
The interaction as a whole -- but especially with Bubbles -- was described as satisfying, with P13 noting, ``It was simply fun to watch!''

\section{Discussion}
We discuss the bigger picture emerging from our results, followed by specific aspects of interest.

\subsection{Word Bubbles Support Continuous Touch Control of Text Generation}\label{sec:discussion_bubbles} %
Our ``Bubbles'' visualisation with its text length and word indicators consistently outperformed \revision{the alternatives}, allowing participants to complete tasks more quickly: 
It averaged \secs{14.71} per task, compared to NoVis (\secs{16.58}) and Lines (\secs{16.50}). 
This speed advantage was evident across tasks with text extension, shortening, and combinations.
Bubbles also received the highest usability rating (SUS: 85.54) and the lowest perceived workload (NASA-TLX score: 1.98).

In addition, participants reported that Bubbles made interactions feel smooth and natural, with many describing the gestures as intuitive and engaging. 
The visual separation of words, sentences, and deleted content provided by Bubbles enhanced the sense of control, making it the preferred method. %

\subsection{Direct Interaction vs Conversations}
When comparing our gestures to the typical conversational LLM interface, participants rated the touch gestures as more usable (81 vs 52.5 SUS score) and less mentally taxing (2.06 vs 3.15 NASA-TLX). 
This is further reflected in %
all subjective measures, including satisfaction, ease of use, control, and efficiency.
Indeed, using gestures led to \pct{58} reduced task times (\secs{56.35} vs \secs{134.86}).

We conclude that our gesture-based concept was well-received and demonstrated clear advantages over a chatbot UI. Combined with our visual feedback, it thus presents a promising alternative for future text editing applications.

\subsection{Direct Interaction and Authorship}
With gestures, participants generally felt more control over text length and the generation process (\cref{fig:own_likert_exp2}, \cref{sec:ss_interview_perception}) yet many did not perceive themselves as the authors of this text. 
This is unsurprising, as \revision{our study} (a) provided given text, and (b) focused more on adjusting text length than altering content.

With our prototype, users could modify tone and input custom prompts with a long press (\cref{fig:long_press}). 
We expected the ChatGPT-like app to receive similar or even higher ratings in the ``authorship'' category, given that participants entered prompts themselves. 
However, this was not reflected in their ratings -- gestures scored higher. 
We hypothesize that the expected fall off in perceived authorship was mitigated due to participants feeling more like authors when they did not need to switch contexts or interact with an external (chatbot) app. %
This effect might become even stronger as more LLM capabilities are controlled through direct (touch) interaction.

\revision{More broadly,} previous research has shown that perceived authorship does not necessarily equate to actually \revision{having entered the text} \cite{AIghostwriter2024}. 
\revision{Thus, as HCI research pursues} interaction with AI \revision{to become} increasingly direct and seamless, %
we believe it is important to also take a nuanced look at the concept of authorship \revision{and the interaction method's impact on its perception}.

\subsection{Direct Interaction and Interaction Metaphors}
When using conversational UI elements, some users tend to perceive the AI more as a writing partner \cite{diarymateKim2024, benharrak2024aipersonas}, rather than a tool. %
Direct interaction has the potential to change this: 
Controlling the LLM directly feels more natural and gives a greater sense of control, as reflected in our findings. 
It also removes the mental association with a writing partner, as there’s no conversational interface -- just the user's own movements. 
As LLM interactions become more seamless, users might begin to view the LLM as part of their cognitive process \cite{gptMeMeshi2024, bhat2023suggestionmodel}, \revision{which could be explored in the} future.

\subsection{Impact of the Visual Feedback on User Interaction and Cognitive Load}
The choice of visual feedback had a significant impact on how users interact with and perceive the system.

\revision{Beyond the benefits of ``Bubbles'' discussed above (\cref{sec:discussion_bubbles})}, this design also impacted gesture execution: 
With Bubbles, participants approached the intended text length more consistently and with less overshooting. 
\revision{Without} visual feedback, \revision{they instead} had to wait for all words to be generated before gauging text length, as both the length and content were displayed through the same medium -- words. 
This presents a unique challenge for interaction with LLMs, where generation is faster than reading, but slower than users can assess the intended length.

Bubbles decoupled text length indication from actual words, allowing users to gauge the amount of text and start reading sooner. We hypothesize that differences in execution patterns stem from this decoupling. 

As less overshooting occurred \revision{with Bubbles}, people had to read less generated text, which may have lowered cognitive load. 
This could explain the reduced mental demand during writing when using Bubbles, as measured by the NASA-TLX (\cref{fig:own_likert_exp2}). 

Reading, attention, and related indicators of cognitive demand were not measured beyond self-reports (e.g. with eye tracking), which could be a focus for future research.

\subsection{Exploring the Design Space of Mobile Touch Interaction with LLMs}
After defining the design space, we adopted a depth-first approach in this paper and focused on exploring one specific LLM capability and gesture in detail: 
\spread{}, along with the inverse, \pinch{}. 

In future studies, we plan to explore additional alternatives and command mappings:
\begin{itemize}
    \itemsep 2mm
    \item \textit{Spread-to-Elaborate, Pinch-to-Summarize:} Currently, spreading appends text at the end of a sentence. Future versions could use spreading over existing \revision{paragraphs to extend them by inserting} elaborations. Similarly, pinching could trigger AI-generated summaries.
    \item \textit{Swipe-to-Rephrase, Rotate-for-Tone:} Swiping over text passages could prompt the LLM to rephrase them or generate synonyms for selected words, while a \revision{rotation} gesture could adjust tone (e.g. ``dialling up/down'' formality).
    \item \textit{Tap-to-Talk:} Special gestures such as a triple tap or a three-finger swipe could request detailed explanations or feedback from the LLM.
    \item \textit{Other Input Modalities:} Beyond touch-based interaction, shaking the phone could trigger rewording and tilting could provide finer control. %
\end{itemize}

\subsection{Limitations}
Our study comes with limitations.
Our prototype was limited to two gestures, within a simple text field, and tested on one device. 
Integrating our prototyped functionality into a larger mobile application (e.g. respecting existing gestures) was beyond our scope.

We covered an initial sample with diversity in some demographics (age, technology use). 
Future work should further evaluate the gestures with a larger, more diverse population.

Participants preferred touch controls over a conversational UI for the tested tasks of text generation and shortening. 
This should not be generalised as an overall preference. 
It is likely that conversational UIs, which offer much more open functionality than gestures, would still be preferred in other tasks (e.g. complex information retrieval, interactive dialogues).
We plan to expand our prototype and compare interactions across further tasks in the future.

\section{Conclusion}

We have introduced a design space for mobile touch interaction with generative AI and invite the community to join us in exploring it further. %
Our own empirical investigation in this paper focused on LLM-supported text length modifications using the \spread{} and \pinch{} gestures.

We developed a novel visual feedback loop with ``word bubbles'' to handle the new requirements posed by mapping continuous touch gestures to continuous text generation. In particular, our design handles both irregular lag from the LLM and provides continuous gesture execution feedback. 
We implemented a functional prototype and evaluated it in two experiments.

Participants consistently performed tasks faster with the Bubble design and found it more intuitive, enjoyable, and easier to use, with a lower perceived workload. 
They preferred the gesture-based interaction for its natural feel and enhanced control. 
The usability of the system was rated higher, and tasks were completed significantly faster with touch gestures, compared to a conversational UI in the style of ChatGPT. %

These results highlight the potential of gesture-based controls for mobile text editing, and more direct interaction techniques for generative AI in general.

Current design trends often introduce conversational UI views and elements for generative AI. We are excited about the alternative potentials to be explored in an emerging research direction (also see \cite{directGPT, Chung2022taleBrush, Chung2024patchview}) that seeks to enable direct manipulation principles for the new capabilities of LLMs.

We release our project material in this repository to facilitate future research:

\url{https://osf.io/h3wyf/}

\begin{acks}
Funded by the Deutsche Forschungsgemeinschaft (DFG, German Research Foundation) -- 525037874.
This project is funded by the Bavarian State Ministry of Science and the Arts and coordinated by the Bavarian Research Institute for Digital Transformation (bidt).
\end{acks}

\bibliographystyle{ACM-Reference-Format}
\bibliography{bibliography}


\begin{thebibliography}{39}


\ifx \showCODEN    \undefined \def \showCODEN     #1{\unskip}     \fi
\ifx \showDOI      \undefined \def \showDOI       #1{#1}\fi
\ifx \showISBNx    \undefined \def \showISBNx     #1{\unskip}     \fi
\ifx \showISBNxiii \undefined \def \showISBNxiii  #1{\unskip}     \fi
\ifx \showISSN     \undefined \def \showISSN      #1{\unskip}     \fi
\ifx \showLCCN     \undefined \def \showLCCN      #1{\unskip}     \fi
\ifx \shownote     \undefined \def \shownote      #1{#1}          \fi
\ifx \showarticletitle \undefined \def \showarticletitle #1{#1}   \fi
\ifx \showURL      \undefined \def \showURL       {\relax}        \fi
\providecommand\bibfield[2]{#2}
\providecommand\bibinfo[2]{#2}
\providecommand\natexlab[1]{#1}
\providecommand\showeprint[2][]{arXiv:#2}

\bibitem[Alvina et~al\mbox{.}(2017)]%
        {Alvina2017commandboard}
\bibfield{author}{\bibinfo{person}{Jessalyn Alvina}, \bibinfo{person}{Carla~F.
  Griggio}, \bibinfo{person}{Xiaojun Bi}, {and} \bibinfo{person}{Wendy~E.
  Mackay}.} \bibinfo{year}{2017}\natexlab{}.
\newblock \showarticletitle{CommandBoard: Creating a General-Purpose Command
  Gesture Input Space for Soft Keyboard}. In
  \bibinfo{booktitle}{\emph{Proceedings of the 30th Annual ACM Symposium on
  User Interface Software and Technology}} (Qu\'{e}bec City, QC, Canada)
  \emph{(\bibinfo{series}{UIST '17})}. \bibinfo{publisher}{Association for
  Computing Machinery}, \bibinfo{address}{New York, NY, USA},
  \bibinfo{pages}{17–28}.
\newblock
\showISBNx{9781450349819}
\urldef\tempurl%
\url{https://doi.org/10.1145/3126594.3126639}
\showDOI{\tempurl}


\bibitem[Bangor et~al\mbox{.}(2009)]%
        {susBangor2009}
\bibfield{author}{\bibinfo{person}{Aaron Bangor}, \bibinfo{person}{Philip
  Kortum}, {and} \bibinfo{person}{James~T. Miller}.}
  \bibinfo{year}{2009}\natexlab{}.
\newblock \showarticletitle{Determining what individual SUS scores mean: Adding
  an adjective rating scale}.
\newblock \bibinfo{journal}{\emph{Journal of Usability Studies}}
  \bibinfo{volume}{4}, \bibinfo{number}{3} (\bibinfo{year}{2009}),
  \bibinfo{pages}{114--123}.
\newblock


\bibitem[Banovic et~al\mbox{.}(2019)]%
        {Banovic2019}
\bibfield{author}{\bibinfo{person}{Nikola Banovic}, \bibinfo{person}{Ticha
  Sethapakdi}, \bibinfo{person}{Yasasvi Hari}, \bibinfo{person}{Anind~K. Dey},
  {and} \bibinfo{person}{Jennifer Mankoff}.} \bibinfo{year}{2019}\natexlab{}.
\newblock \showarticletitle{The Limits of Expert Text Entry Speed on Mobile
  Keyboards with Autocorrect}. In \bibinfo{booktitle}{\emph{Proceedings of the
  21st International Conference on Human-Computer Interaction with Mobile
  Devices and Services}} (Taipei, Taiwan) \emph{(\bibinfo{series}{MobileHCI
  '19})}. \bibinfo{publisher}{Association for Computing Machinery},
  \bibinfo{address}{New York, NY, USA}, Article \bibinfo{articleno}{15},
  \bibinfo{numpages}{12}~pages.
\newblock
\showISBNx{9781450368254}
\urldef\tempurl%
\url{https://doi.org/10.1145/3338286.3340126}
\showDOI{\tempurl}


\bibitem[Bates et~al\mbox{.}(2015)]%
        {Bates2015}
\bibfield{author}{\bibinfo{person}{Douglas Bates}, \bibinfo{person}{Martin
  M{\"a}chler}, \bibinfo{person}{Ben Bolker}, {and} \bibinfo{person}{Steve
  Walker}.} \bibinfo{year}{2015}\natexlab{}.
\newblock \showarticletitle{Fitting Linear Mixed-Effects Models Using {lme4}}.
\newblock \bibinfo{journal}{\emph{Journal of Statistical Software}}
  \bibinfo{volume}{67}, \bibinfo{number}{1} (\bibinfo{year}{2015}),
  \bibinfo{pages}{1--48}.
\newblock
\urldef\tempurl%
\url{https://doi.org/10.18637/jss.v067.i01}
\showDOI{\tempurl}


\bibitem[Beaudouin-Lafon(2000)]%
        {BeaudouinLafon2000instrumental}
\bibfield{author}{\bibinfo{person}{Michel Beaudouin-Lafon}.}
  \bibinfo{year}{2000}\natexlab{}.
\newblock \showarticletitle{Instrumental interaction: an interaction model for
  designing post-WIMP user interfaces}. In
  \bibinfo{booktitle}{\emph{Proceedings of the SIGCHI Conference on Human
  Factors in Computing Systems}} (The Hague, The Netherlands)
  \emph{(\bibinfo{series}{CHI '00})}. \bibinfo{publisher}{Association for
  Computing Machinery}, \bibinfo{address}{New York, NY, USA},
  \bibinfo{pages}{446–453}.
\newblock
\showISBNx{1581132166}
\urldef\tempurl%
\url{https://doi.org/10.1145/332040.332473}
\showDOI{\tempurl}


\bibitem[Benharrak et~al\mbox{.}(2024)]%
        {benharrak2024aipersonas}
\bibfield{author}{\bibinfo{person}{Karim Benharrak}, \bibinfo{person}{Tim
  Zindulka}, \bibinfo{person}{Florian Lehmann}, \bibinfo{person}{Hendrik
  Heuer}, {and} \bibinfo{person}{Daniel Buschek}.}
  \bibinfo{year}{2024}\natexlab{}.
\newblock \showarticletitle{Writer-Defined AI Personas for On-Demand Feedback
  Generation}. In \bibinfo{booktitle}{\emph{Proceedings of the CHI Conference
  on Human Factors in Computing Systems}} (Honolulu, HI, USA)
  \emph{(\bibinfo{series}{CHI '24})}. \bibinfo{publisher}{Association for
  Computing Machinery}, \bibinfo{address}{New York, NY, USA}, Article
  \bibinfo{articleno}{1049}, \bibinfo{numpages}{18}~pages.
\newblock
\showISBNx{9798400703300}
\urldef\tempurl%
\url{https://doi.org/10.1145/3613904.3642406}
\showDOI{\tempurl}


\bibitem[Bhat et~al\mbox{.}(2023)]%
        {bhat2023suggestionmodel}
\bibfield{author}{\bibinfo{person}{Advait Bhat}, \bibinfo{person}{Saaket
  Agashe}, \bibinfo{person}{Parth Oberoi}, \bibinfo{person}{Niharika Mohile},
  \bibinfo{person}{Ravi Jangir}, {and} \bibinfo{person}{Anirudha Joshi}.}
  \bibinfo{year}{2023}\natexlab{}.
\newblock \showarticletitle{Interacting with Next-Phrase Suggestions: How
  Suggestion Systems Aid and Influence the Cognitive Processes of Writing}. In
  \bibinfo{booktitle}{\emph{Proceedings of the 28th International Conference on
  Intelligent User Interfaces}} (Sydney, NSW, Australia)
  \emph{(\bibinfo{series}{IUI '23})}. \bibinfo{publisher}{Association for
  Computing Machinery}, \bibinfo{address}{New York, NY, USA},
  \bibinfo{pages}{436–452}.
\newblock
\showISBNx{9798400701061}
\urldef\tempurl%
\url{https://doi.org/10.1145/3581641.3584060}
\showDOI{\tempurl}


\bibitem[Buschek(2024)]%
        {Buschek2024collage}
\bibfield{author}{\bibinfo{person}{Daniel Buschek}.}
  \bibinfo{year}{2024}\natexlab{}.
\newblock \showarticletitle{Collage is the New Writing: Exploring the
  Fragmentation of Text and User Interfaces in AI Tools}. In
  \bibinfo{booktitle}{\emph{Proceedings of the 2024 ACM Designing Interactive
  Systems Conference}} (Copenhagen, Denmark) \emph{(\bibinfo{series}{DIS
  '24})}. \bibinfo{publisher}{Association for Computing Machinery},
  \bibinfo{address}{New York, NY, USA}, \bibinfo{pages}{2719–2737}.
\newblock
\showISBNx{9798400705830}
\urldef\tempurl%
\url{https://doi.org/10.1145/3643834.3660681}
\showDOI{\tempurl}


\bibitem[Chung et~al\mbox{.}(2022)]%
        {Chung2022taleBrush}
\bibfield{author}{\bibinfo{person}{John Joon~Young Chung},
  \bibinfo{person}{Wooseok Kim}, \bibinfo{person}{Kang~Min Yoo},
  \bibinfo{person}{Hwaran Lee}, \bibinfo{person}{Eytan Adar}, {and}
  \bibinfo{person}{Minsuk Chang}.} \bibinfo{year}{2022}\natexlab{}.
\newblock \showarticletitle{TaleBrush: Sketching Stories with Generative
  Pretrained Language Models}. In \bibinfo{booktitle}{\emph{Proceedings of the
  2022 CHI Conference on Human Factors in Computing Systems}} (New Orleans, LA,
  USA) \emph{(\bibinfo{series}{CHI '22})}. \bibinfo{publisher}{Association for
  Computing Machinery}, \bibinfo{address}{New York, NY, USA}, Article
  \bibinfo{articleno}{209}, \bibinfo{numpages}{19}~pages.
\newblock
\showISBNx{9781450391573}
\urldef\tempurl%
\url{https://doi.org/10.1145/3491102.3501819}
\showDOI{\tempurl}


\bibitem[Chung and Kreminski(2024)]%
        {Chung2024patchview}
\bibfield{author}{\bibinfo{person}{John Joon~Young Chung} {and}
  \bibinfo{person}{Max Kreminski}.} \bibinfo{year}{2024}\natexlab{}.
\newblock \bibinfo{title}{Patchview: {LLM}-{Powered} {Worldbuilding} with
  {Generative} {Dust} and {Magnet} {Visualization}}.
\newblock
\newblock
\urldef\tempurl%
\url{https://doi.org/10.1145/3654777.3676352}
\showDOI{\tempurl}
\newblock
\shownote{arXiv:2408.04112 [cs]}.


\bibitem[Cooper et~al\mbox{.}(2014)]%
        {Cooper2014aboutface}
\bibfield{author}{\bibinfo{person}{Alan Cooper}, \bibinfo{person}{Robert
  Reimann}, \bibinfo{person}{David Cronin}, {and} \bibinfo{person}{Christopher
  Noessel}.} \bibinfo{year}{2014}\natexlab{}.
\newblock \bibinfo{booktitle}{\emph{About face: the essentials of interaction
  design}}.
\newblock \bibinfo{publisher}{John Wiley \& Sons}.
\newblock


\bibitem[Cui et~al\mbox{.}(2020)]%
        {Cui2020justcorrect}
\bibfield{author}{\bibinfo{person}{Wenzhe Cui}, \bibinfo{person}{Suwen Zhu},
  \bibinfo{person}{Mingrui~Ray Zhang}, \bibinfo{person}{H.~Andrew Schwartz},
  \bibinfo{person}{Jacob~O. Wobbrock}, {and} \bibinfo{person}{Xiaojun Bi}.}
  \bibinfo{year}{2020}\natexlab{}.
\newblock \showarticletitle{JustCorrect: Intelligent Post Hoc Text Correction
  Techniques on Smartphones}. In \bibinfo{booktitle}{\emph{Proceedings of the
  33rd Annual ACM Symposium on User Interface Software and Technology}}
  (Virtual Event, USA) \emph{(\bibinfo{series}{UIST '20})}.
  \bibinfo{publisher}{Association for Computing Machinery},
  \bibinfo{address}{New York, NY, USA}, \bibinfo{pages}{487–499}.
\newblock
\showISBNx{9781450375146}
\urldef\tempurl%
\url{https://doi.org/10.1145/3379337.3415857}
\showDOI{\tempurl}


\bibitem[Dang et~al\mbox{.}(2023)]%
        {Dang2023choice}
\bibfield{author}{\bibinfo{person}{Hai Dang}, \bibinfo{person}{Sven Goller},
  \bibinfo{person}{Florian Lehmann}, {and} \bibinfo{person}{Daniel Buschek}.}
  \bibinfo{year}{2023}\natexlab{}.
\newblock \showarticletitle{Choice Over Control: How Users Write with Large
  Language Models using Diegetic and Non-Diegetic Prompting}. In
  \bibinfo{booktitle}{\emph{Proceedings of the 2023 CHI Conference on Human
  Factors in Computing Systems}} (Hamburg, Germany) \emph{(\bibinfo{series}{CHI
  '23})}. \bibinfo{publisher}{Association for Computing Machinery},
  \bibinfo{address}{New York, NY, USA}, Article \bibinfo{articleno}{408},
  \bibinfo{numpages}{17}~pages.
\newblock
\showISBNx{9781450394215}
\urldef\tempurl%
\url{https://doi.org/10.1145/3544548.3580969}
\showDOI{\tempurl}


\bibitem[Draxler et~al\mbox{.}(2024)]%
        {AIghostwriter2024}
\bibfield{author}{\bibinfo{person}{Fiona Draxler}, \bibinfo{person}{Anna
  Werner}, \bibinfo{person}{Florian Lehmann}, \bibinfo{person}{Matthias Hoppe},
  \bibinfo{person}{Albrecht Schmidt}, \bibinfo{person}{Daniel Buschek}, {and}
  \bibinfo{person}{Robin Welsch}.} \bibinfo{year}{2024}\natexlab{}.
\newblock \showarticletitle{The AI Ghostwriter Effect: When Users do not
  Perceive Ownership of AI-Generated Text but Self-Declare as Authors}.
\newblock \bibinfo{journal}{\emph{ACM Trans. Comput.-Hum. Interact.}}
  \bibinfo{volume}{31}, \bibinfo{number}{2}, Article \bibinfo{articleno}{25}
  (\bibinfo{date}{feb} \bibinfo{year}{2024}), \bibinfo{numpages}{40}~pages.
\newblock
\showISSN{1073-0516}
\urldef\tempurl%
\url{https://doi.org/10.1145/3637875}
\showDOI{\tempurl}


\bibitem[Fu et~al\mbox{.}(2024)]%
        {Fu2024texttoself}
\bibfield{author}{\bibinfo{person}{Yue Fu}, \bibinfo{person}{Sami Foell},
  \bibinfo{person}{Xuhai Xu}, {and} \bibinfo{person}{Alexis Hiniker}.}
  \bibinfo{year}{2024}\natexlab{}.
\newblock \showarticletitle{From Text to Self: Users’ Perception of AIMC
  Tools on Interpersonal Communication and Self}. In
  \bibinfo{booktitle}{\emph{Proceedings of the CHI Conference on Human Factors
  in Computing Systems}} (Honolulu, HI, USA) \emph{(\bibinfo{series}{CHI
  '24})}. \bibinfo{publisher}{Association for Computing Machinery},
  \bibinfo{address}{New York, NY, USA}, Article \bibinfo{articleno}{977},
  \bibinfo{numpages}{17}~pages.
\newblock
\showISBNx{9798400703300}
\urldef\tempurl%
\url{https://doi.org/10.1145/3613904.3641955}
\showDOI{\tempurl}


\bibitem[Hartson(2012)]%
        {hartson2012ux}
\bibfield{author}{\bibinfo{person}{Rex Hartson}.}
  \bibinfo{year}{2012}\natexlab{}.
\newblock \bibinfo{booktitle}{\emph{The UX Book: Process and Guidelines for
  Ensuring a Quality User Experience}}.
\newblock \bibinfo{publisher}{Elsevier}.
\newblock


\bibitem[Ishii and Shizuki(2016)]%
        {Ishii2016callout}
\bibfield{author}{\bibinfo{person}{Akira Ishii} {and} \bibinfo{person}{Buntarou
  Shizuki}.} \bibinfo{year}{2016}\natexlab{}.
\newblock \showarticletitle{Exploring callout design in selection task for
  ultra-small touch screen devices}. In \bibinfo{booktitle}{\emph{Proceedings
  of the 28th Australian Conference on Computer-Human Interaction}}
  (Launceston, Tasmania, Australia) \emph{(\bibinfo{series}{OzCHI '16})}.
  \bibinfo{publisher}{Association for Computing Machinery},
  \bibinfo{address}{New York, NY, USA}, \bibinfo{pages}{426–434}.
\newblock
\showISBNx{9781450346184}
\urldef\tempurl%
\url{https://doi.org/10.1145/3010915.3010922}
\showDOI{\tempurl}


\bibitem[Kim et~al\mbox{.}(2024)]%
        {diarymateKim2024}
\bibfield{author}{\bibinfo{person}{Taewan Kim}, \bibinfo{person}{Donghoon
  Shin}, \bibinfo{person}{Young-Ho Kim}, {and} \bibinfo{person}{Hwajung Hong}.}
  \bibinfo{year}{2024}\natexlab{}.
\newblock \showarticletitle{DiaryMate: Understanding User Perceptions and
  Experience in Human-AI Collaboration for Personal Journaling}. In
  \bibinfo{booktitle}{\emph{Proceedings of the CHI Conference on Human Factors
  in Computing Systems}} (Honolulu, HI, USA) \emph{(\bibinfo{series}{CHI
  '24})}. \bibinfo{publisher}{Association for Computing Machinery},
  \bibinfo{address}{New York, NY, USA}, Article \bibinfo{articleno}{1046},
  \bibinfo{numpages}{15}~pages.
\newblock
\showISBNx{9798400703300}
\urldef\tempurl%
\url{https://doi.org/10.1145/3613904.3642693}
\showDOI{\tempurl}


\bibitem[Kristensson and Vertanen(2014)]%
        {Kristensson2014inviscid}
\bibfield{author}{\bibinfo{person}{Per~Ola Kristensson} {and}
  \bibinfo{person}{Keith Vertanen}.} \bibinfo{year}{2014}\natexlab{}.
\newblock \showarticletitle{The inviscid text entry rate and its application as
  a grand goal for mobile text entry}. In \bibinfo{booktitle}{\emph{Proceedings
  of the 16th International Conference on Human-Computer Interaction with
  Mobile Devices \& Services}} (Toronto, ON, Canada)
  \emph{(\bibinfo{series}{MobileHCI '14})}. \bibinfo{publisher}{Association for
  Computing Machinery}, \bibinfo{address}{New York, NY, USA},
  \bibinfo{pages}{335–338}.
\newblock
\showISBNx{9781450330046}
\urldef\tempurl%
\url{https://doi.org/10.1145/2628363.2628405}
\showDOI{\tempurl}


\bibitem[Kurtenbach(1993)]%
        {Kurtenbach1993markingmenus}
\bibfield{author}{\bibinfo{person}{Gordon Kurtenbach}.}
  \bibinfo{year}{1993}\natexlab{}.
\newblock \bibinfo{booktitle}{\emph{The design and evaluation of marking
  menus.}}
\newblock \bibinfo{publisher}{University of Toronto Toronto}.
\newblock


\bibitem[Kurtenbach and Buxton(1994)]%
        {Kurtenbach1994markingmenus}
\bibfield{author}{\bibinfo{person}{Gordon Kurtenbach} {and}
  \bibinfo{person}{William Buxton}.} \bibinfo{year}{1994}\natexlab{}.
\newblock \showarticletitle{User learning and performance with marking menus}.
  In \bibinfo{booktitle}{\emph{Proceedings of the SIGCHI Conference on Human
  Factors in Computing Systems}} (Boston, Massachusetts, USA)
  \emph{(\bibinfo{series}{CHI '94})}. \bibinfo{publisher}{Association for
  Computing Machinery}, \bibinfo{address}{New York, NY, USA},
  \bibinfo{pages}{258–264}.
\newblock
\showISBNx{0897916506}
\urldef\tempurl%
\url{https://doi.org/10.1145/191666.191759}
\showDOI{\tempurl}


\bibitem[Kuznetsova et~al\mbox{.}(2017)]%
        {Kuznetsova2017}
\bibfield{author}{\bibinfo{person}{Alexandra Kuznetsova},
  \bibinfo{person}{Per~B. Brockhoff}, {and} \bibinfo{person}{Rune H.~B.
  Christensen}.} \bibinfo{year}{2017}\natexlab{}.
\newblock \showarticletitle{{lmerTest} Package: Tests in Linear Mixed Effects
  Models}.
\newblock \bibinfo{journal}{\emph{Journal of Statistical Software}}
  \bibinfo{volume}{82}, \bibinfo{number}{13} (\bibinfo{year}{2017}),
  \bibinfo{pages}{1--26}.
\newblock
\urldef\tempurl%
\url{https://doi.org/10.18637/jss.v082.i13}
\showDOI{\tempurl}


\bibitem[Lee et~al\mbox{.}(2024)]%
        {Lee2024dsiiwa}
\bibfield{author}{\bibinfo{person}{Mina Lee}, \bibinfo{person}{Katy~Ilonka
  Gero}, \bibinfo{person}{John Joon~Young Chung},
  \bibinfo{person}{Simon~Buckingham Shum}, \bibinfo{person}{Vipul Raheja},
  \bibinfo{person}{Hua Shen}, \bibinfo{person}{Subhashini Venugopalan},
  \bibinfo{person}{Thiemo Wambsganss}, \bibinfo{person}{David Zhou},
  \bibinfo{person}{Emad~A. Alghamdi}, \bibinfo{person}{Tal August},
  \bibinfo{person}{Avinash Bhat}, \bibinfo{person}{Madiha~Zahrah Choksi},
  \bibinfo{person}{Senjuti Dutta}, \bibinfo{person}{Jin~L.C. Guo},
  \bibinfo{person}{Md~Naimul Hoque}, \bibinfo{person}{Yewon Kim},
  \bibinfo{person}{Simon Knight}, \bibinfo{person}{Seyed~Parsa Neshaei},
  \bibinfo{person}{Antonette Shibani}, \bibinfo{person}{Disha Shrivastava},
  \bibinfo{person}{Lila Shroff}, \bibinfo{person}{Agnia Sergeyuk},
  \bibinfo{person}{Jessi Stark}, \bibinfo{person}{Sarah Sterman},
  \bibinfo{person}{Sitong Wang}, \bibinfo{person}{Antoine Bosselut},
  \bibinfo{person}{Daniel Buschek}, \bibinfo{person}{Joseph~Chee Chang},
  \bibinfo{person}{Sherol Chen}, \bibinfo{person}{Max Kreminski},
  \bibinfo{person}{Joonsuk Park}, \bibinfo{person}{Roy Pea},
  \bibinfo{person}{Eugenia Ha~Rim Rho}, \bibinfo{person}{Zejiang Shen}, {and}
  \bibinfo{person}{Pao Siangliulue}.} \bibinfo{year}{2024}\natexlab{}.
\newblock \showarticletitle{A Design Space for Intelligent and Interactive
  Writing Assistants}. In \bibinfo{booktitle}{\emph{Proceedings of the CHI
  Conference on Human Factors in Computing Systems}} (Honolulu, HI, USA)
  \emph{(\bibinfo{series}{CHI '24})}. \bibinfo{publisher}{Association for
  Computing Machinery}, \bibinfo{address}{New York, NY, USA}, Article
  \bibinfo{articleno}{1054}, \bibinfo{numpages}{35}~pages.
\newblock
\showISBNx{9798400703300}
\urldef\tempurl%
\url{https://doi.org/10.1145/3613904.3642697}
\showDOI{\tempurl}


\bibitem[Lehmann et~al\mbox{.}(2022)]%
        {Lehmann2022suggVsCont}
\bibfield{author}{\bibinfo{person}{Florian Lehmann}, \bibinfo{person}{Niklas
  Markert}, \bibinfo{person}{Hai Dang}, {and} \bibinfo{person}{Daniel
  Buschek}.} \bibinfo{year}{2022}\natexlab{}.
\newblock \showarticletitle{Suggestion Lists vs. Continuous Generation:
  Interaction Design for Writing with Generative Models on Mobile Devices
  Affect Text Length, Wording and Perceived Authorship}. In
  \bibinfo{booktitle}{\emph{Proceedings of Mensch Und Computer 2022}}
  (Darmstadt, Germany) \emph{(\bibinfo{series}{MuC '22})}.
  \bibinfo{publisher}{Association for Computing Machinery},
  \bibinfo{address}{New York, NY, USA}, \bibinfo{pages}{192–208}.
\newblock
\showISBNx{9781450396905}
\urldef\tempurl%
\url{https://doi.org/10.1145/3543758.3543947}
\showDOI{\tempurl}


\bibitem[Liu et~al\mbox{.}(2022)]%
        {Liu2022aimailperception}
\bibfield{author}{\bibinfo{person}{Yihe Liu}, \bibinfo{person}{Anushk Mittal},
  \bibinfo{person}{Diyi Yang}, {and} \bibinfo{person}{Amy Bruckman}.}
  \bibinfo{year}{2022}\natexlab{}.
\newblock \showarticletitle{Will AI Console Me when I Lose my Pet?
  Understanding Perceptions of AI-Mediated Email Writing}. In
  \bibinfo{booktitle}{\emph{Proceedings of the 2022 CHI Conference on Human
  Factors in Computing Systems}} (New Orleans, LA, USA)
  \emph{(\bibinfo{series}{CHI '22})}. \bibinfo{publisher}{Association for
  Computing Machinery}, \bibinfo{address}{New York, NY, USA}, Article
  \bibinfo{articleno}{474}, \bibinfo{numpages}{13}~pages.
\newblock
\showISBNx{9781450391573}
\urldef\tempurl%
\url{https://doi.org/10.1145/3491102.3517731}
\showDOI{\tempurl}


\bibitem[Malloch et~al\mbox{.}(2017)]%
        {Malloch2017fieldpathward}
\bibfield{author}{\bibinfo{person}{Joseph Malloch}, \bibinfo{person}{Carla~F.
  Griggio}, \bibinfo{person}{Joanna McGrenere}, {and} \bibinfo{person}{Wendy~E.
  Mackay}.} \bibinfo{year}{2017}\natexlab{}.
\newblock \showarticletitle{Fieldward and Pathward: Dynamic Guides for Defining
  Your Own Gestures}. In \bibinfo{booktitle}{\emph{Proceedings of the 2017 CHI
  Conference on Human Factors in Computing Systems}} (Denver, Colorado, USA)
  \emph{(\bibinfo{series}{CHI '17})}. \bibinfo{publisher}{Association for
  Computing Machinery}, \bibinfo{address}{New York, NY, USA},
  \bibinfo{pages}{4266–4277}.
\newblock
\showISBNx{9781450346559}
\urldef\tempurl%
\url{https://doi.org/10.1145/3025453.3025764}
\showDOI{\tempurl}


\bibitem[Masson et~al\mbox{.}(2024)]%
        {directGPT}
\bibfield{author}{\bibinfo{person}{Damien Masson}, \bibinfo{person}{Sylvain
  Malacria}, \bibinfo{person}{G\'{e}ry Casiez}, {and} \bibinfo{person}{Daniel
  Vogel}.} \bibinfo{year}{2024}\natexlab{}.
\newblock \showarticletitle{DirectGPT: A Direct Manipulation Interface to
  Interact with Large Language Models}. In
  \bibinfo{booktitle}{\emph{Proceedings of the CHI Conference on Human Factors
  in Computing Systems}} (Honolulu, HI, USA) \emph{(\bibinfo{series}{CHI
  '24})}. \bibinfo{publisher}{Association for Computing Machinery},
  \bibinfo{address}{New York, NY, USA}, Article \bibinfo{articleno}{975},
  \bibinfo{numpages}{16}~pages.
\newblock
\showISBNx{9798400703300}
\urldef\tempurl%
\url{https://doi.org/10.1145/3613904.3642462}
\showDOI{\tempurl}


\bibitem[Meshi(2024)]%
        {gptMeMeshi2024}
\bibfield{author}{\bibinfo{person}{Avital Meshi}.}
  \bibinfo{year}{2024}\natexlab{}.
\newblock \showarticletitle{GPT-ME: A Human-AI Cognitive Assemblage}.
\newblock \bibinfo{journal}{\emph{Proc. ACM Comput. Graph. Interact. Tech.}}
  \bibinfo{volume}{7}, \bibinfo{number}{4}, Article \bibinfo{articleno}{55}
  (\bibinfo{date}{jul} \bibinfo{year}{2024}), \bibinfo{numpages}{8}~pages.
\newblock
\urldef\tempurl%
\url{https://doi.org/10.1145/3664214}
\showDOI{\tempurl}


\bibitem[M\"{u}ller et~al\mbox{.}(2017)]%
        {mouse_mueller2017}
\bibfield{author}{\bibinfo{person}{J\"{o}rg M\"{u}ller}, \bibinfo{person}{Antti
  Oulasvirta}, {and} \bibinfo{person}{Roderick Murray-Smith}.}
  \bibinfo{year}{2017}\natexlab{}.
\newblock \showarticletitle{Control Theoretic Models of Pointing}.
\newblock \bibinfo{journal}{\emph{ACM Trans. Comput.-Hum. Interact.}}
  \bibinfo{volume}{24}, \bibinfo{number}{4}, Article \bibinfo{articleno}{27}
  (\bibinfo{date}{aug} \bibinfo{year}{2017}), \bibinfo{numpages}{36}~pages.
\newblock
\showISSN{1073-0516}
\urldef\tempurl%
\url{https://doi.org/10.1145/3121431}
\showDOI{\tempurl}


\bibitem[Palin et~al\mbox{.}(2019)]%
        {Palin2019}
\bibfield{author}{\bibinfo{person}{Kseniia Palin}, \bibinfo{person}{Anna~Maria
  Feit}, \bibinfo{person}{Sunjun Kim}, \bibinfo{person}{Per~Ola Kristensson},
  {and} \bibinfo{person}{Antti Oulasvirta}.} \bibinfo{year}{2019}\natexlab{}.
\newblock \showarticletitle{How do People Type on Mobile Devices? Observations
  from a Study with 37,000 Volunteers}. In
  \bibinfo{booktitle}{\emph{Proceedings of the 21st International Conference on
  Human-Computer Interaction with Mobile Devices and Services}} (Taipei,
  Taiwan) \emph{(\bibinfo{series}{MobileHCI '19})}.
  \bibinfo{publisher}{Association for Computing Machinery},
  \bibinfo{address}{New York, NY, USA}, Article \bibinfo{articleno}{9},
  \bibinfo{numpages}{12}~pages.
\newblock
\showISBNx{9781450368254}
\urldef\tempurl%
\url{https://doi.org/10.1145/3338286.3340120}
\showDOI{\tempurl}


\bibitem[Quinn and Zhai(2016)]%
        {Quinn2016}
\bibfield{author}{\bibinfo{person}{Philip Quinn} {and} \bibinfo{person}{Shumin
  Zhai}.} \bibinfo{year}{2016}\natexlab{}.
\newblock \showarticletitle{A Cost-Benefit Study of Text Entry Suggestion
  Interaction}. In \bibinfo{booktitle}{\emph{Proceedings of the 2016 CHI
  Conference on Human Factors in Computing Systems}} (San Jose, California,
  USA) \emph{(\bibinfo{series}{CHI '16})}. \bibinfo{publisher}{Association for
  Computing Machinery}, \bibinfo{address}{New York, NY, USA},
  \bibinfo{pages}{83–88}.
\newblock
\showISBNx{9781450333627}
\urldef\tempurl%
\url{https://doi.org/10.1145/2858036.2858305}
\showDOI{\tempurl}


\bibitem[{R Core Team}(2020)]%
        {R2020}
\bibfield{author}{\bibinfo{person}{{R Core Team}}.}
  \bibinfo{year}{2020}\natexlab{}.
\newblock \bibinfo{booktitle}{\emph{R: A Language and Environment for
  Statistical Computing}}.
\newblock R Foundation for Statistical Computing, Vienna, Austria.
\newblock
\urldef\tempurl%
\url{https://www.R-project.org}
\showURL{%
\tempurl}


\bibitem[Robertson et~al\mbox{.}(2021)]%
        {Robertson2021cantreply}
\bibfield{author}{\bibinfo{person}{Ronald~E Robertson},
  \bibinfo{person}{Alexandra Olteanu}, \bibinfo{person}{Fernando Diaz},
  \bibinfo{person}{Milad Shokouhi}, {and} \bibinfo{person}{Peter Bailey}.}
  \bibinfo{year}{2021}\natexlab{}.
\newblock \showarticletitle{“I Can’t Reply with That”: Characterizing
  Problematic Email Reply Suggestions}. In
  \bibinfo{booktitle}{\emph{Proceedings of the 2021 CHI Conference on Human
  Factors in Computing Systems}} (Yokohama, Japan) \emph{(\bibinfo{series}{CHI
  '21})}. \bibinfo{publisher}{Association for Computing Machinery},
  \bibinfo{address}{New York, NY, USA}, Article \bibinfo{articleno}{724},
  \bibinfo{numpages}{18}~pages.
\newblock
\showISBNx{9781450380966}
\urldef\tempurl%
\url{https://doi.org/10.1145/3411764.3445557}
\showDOI{\tempurl}


\bibitem[Skryja and Barcik(2024)]%
        {photonics11060540}
\bibfield{author}{\bibinfo{person}{Petr Skryja} {and} \bibinfo{person}{Peter
  Barcik}.} \bibinfo{year}{2024}\natexlab{}.
\newblock \showarticletitle{Optimal Scanning Pattern for Initial Free-Space
  Optical-Link Alignment}.
\newblock \bibinfo{journal}{\emph{Photonics}} \bibinfo{volume}{11},
  \bibinfo{number}{6} (\bibinfo{year}{2024}).
\newblock
\showISSN{2304-6732}
\urldef\tempurl%
\url{https://doi.org/10.3390/photonics11060540}
\showDOI{\tempurl}


\bibitem[Vogel and Baudisch(2007)]%
        {Vogel2007shift}
\bibfield{author}{\bibinfo{person}{Daniel Vogel} {and} \bibinfo{person}{Patrick
  Baudisch}.} \bibinfo{year}{2007}\natexlab{}.
\newblock \showarticletitle{Shift: a technique for operating pen-based
  interfaces using touch}. In \bibinfo{booktitle}{\emph{Proceedings of the
  SIGCHI Conference on Human Factors in Computing Systems}} (San Jose,
  California, USA) \emph{(\bibinfo{series}{CHI '07})}.
  \bibinfo{publisher}{Association for Computing Machinery},
  \bibinfo{address}{New York, NY, USA}, \bibinfo{pages}{657–666}.
\newblock
\showISBNx{9781595935939}
\urldef\tempurl%
\url{https://doi.org/10.1145/1240624.1240727}
\showDOI{\tempurl}


\bibitem[Wolff et~al\mbox{.}(1998)]%
        {wolff1998acting}
\bibfield{author}{\bibinfo{person}{Fr{\'e}d{\'e}ric Wolff},
  \bibinfo{person}{Antonella De~Angeli}, \bibinfo{person}{Laurent Romary},
  {et~al\mbox{.}}} \bibinfo{year}{1998}\natexlab{}.
\newblock \showarticletitle{Acting on a visual world: The role of perception in
  multimodal HCI}. In \bibinfo{booktitle}{\emph{Proceedings of AAAI Workshop on
  Multimodal Representation}}.
\newblock


\bibitem[Yuan et~al\mbox{.}(2022)]%
        {Yuan20222wordcraft}
\bibfield{author}{\bibinfo{person}{Ann Yuan}, \bibinfo{person}{Andy Coenen},
  \bibinfo{person}{Emily Reif}, {and} \bibinfo{person}{Daphne Ippolito}.}
  \bibinfo{year}{2022}\natexlab{}.
\newblock \showarticletitle{Wordcraft: Story Writing With Large Language
  Models}. In \bibinfo{booktitle}{\emph{Proceedings of the 27th International
  Conference on Intelligent User Interfaces}} (Helsinki, Finland)
  \emph{(\bibinfo{series}{IUI '22})}. \bibinfo{publisher}{Association for
  Computing Machinery}, \bibinfo{address}{New York, NY, USA},
  \bibinfo{pages}{841–852}.
\newblock
\showISBNx{9781450391443}
\urldef\tempurl%
\url{https://doi.org/10.1145/3490099.3511105}
\showDOI{\tempurl}


\bibitem[Zhai and Kristensson(2012)]%
        {Zhai2012gesturekbMagazine}
\bibfield{author}{\bibinfo{person}{Shumin Zhai} {and} \bibinfo{person}{Per~Ola
  Kristensson}.} \bibinfo{year}{2012}\natexlab{}.
\newblock \showarticletitle{The word-gesture keyboard: reimagining keyboard
  interaction}.
\newblock \bibinfo{journal}{\emph{Commun. ACM}} \bibinfo{volume}{55},
  \bibinfo{number}{9} (\bibinfo{date}{sep} \bibinfo{year}{2012}),
  \bibinfo{pages}{91–101}.
\newblock
\showISSN{0001-0782}
\urldef\tempurl%
\url{https://doi.org/10.1145/2330667.2330689}
\showDOI{\tempurl}


\bibitem[Zhang et~al\mbox{.}(2019)]%
        {Zhang2019typthencorrect}
\bibfield{author}{\bibinfo{person}{Mingrui~Ray Zhang}, \bibinfo{person}{He
  Wen}, {and} \bibinfo{person}{Jacob~O. Wobbrock}.}
  \bibinfo{year}{2019}\natexlab{}.
\newblock \showarticletitle{Type, Then Correct: Intelligent Text Correction
  Techniques for Mobile Text Entry Using Neural Networks}. In
  \bibinfo{booktitle}{\emph{Proceedings of the 32nd Annual ACM Symposium on
  User Interface Software and Technology}} (New Orleans, LA, USA)
  \emph{(\bibinfo{series}{UIST '19})}. \bibinfo{publisher}{Association for
  Computing Machinery}, \bibinfo{address}{New York, NY, USA},
  \bibinfo{pages}{843–855}.
\newblock
\showISBNx{9781450368162}
\urldef\tempurl%
\url{https://doi.org/10.1145/3332165.3347924}
\showDOI{\tempurl}


\end{thebibliography}

\appendix

\section{Appendix}

This appendix lists our prompting templates, additional tables, texts, and figures, referenced throughout the paper.

\subsection{Prompting Templates}
\label{sec:appendix_prompts}
\subsubsection{Extend} \mbox{}
\label{prompt:extend}
\begin{lstlisting}
You are a helpful assistant that can extend one sentence of a given input at a time.\n
You will get a Paragraph and you should write one sentence to extend it.\n
You should only answer with that generated sentence, NOTHING ELSE!
\end{lstlisting}

\subsubsection{Find Synonym} \mbox{}
\label{prompt:synonym}
\begin{lstlisting}
You are a helpful assistant in a react application that can find synonyms for a given word.\n
You will get a word and you should find a synonym for it.\n
Find at least one synonym, but the more the better.\n
If the given word doesn't have any synonyms, you should return 'NO SYNONYM'\n
Your answer will be parsed into an array of words, so make sure to return your answer in this style: 'Synonym1, Synonym2, Synonym3' and so on!\n
You should only answer with the formated synonyms, NOTHING ELSE!
\end{lstlisting}

\subsubsection{Find Custom Sentence} \mbox{}
\label{prompt:custom_sentence}
\begin{lstlisting}
You are an AI tasked with transforming user-provided sentences according to their specific instructions. Please follow the guidelines below for each request:\n
 1. Input Format:\n
    - You will find the string '*sentence:*', this is the original sentence provided by the user!\n
    - You will find the string '*prompt:*', this describes how the user wants the sentence to be modified or altered!\n
 2. Transformation Instructions:\n
    - Analyze the provided Sentence and apply the modifications described in the '*prompt*' section to the best of your ability.\n
    - If the requested transformation cannot be accurately performed, respond with the original sentence in section '*sentence:*' without any modifications.\n
    - Ensure that your response contains only the transformed sentence or the original sentence if transformation is not feasible. Do not include any additional text or information!\n
 3. Answer Format:\n
    - Return only the modified sentence or the original sentence if the modification is not possible. Do not include any extra comments, explanations, or additional content.\n
 4. Example:\n
    - User Request: '*sentence:* I will call you tomorrow. *prompt:* Make it sound more polite.'\n
    - Your Response could be: 'I would be happy to call you tomorrow.'\n\n
If you have any difficulty performing the requested transformation, simply return the original sentence in section '*sentence*:' as it is.
\end{lstlisting}

\subsubsection{Rewrite Sentence} \mbox{}
\label{prompt:rewrite_sentence}
\begin{lstlisting}
You are a helpful assistant in a react application that should rewrite a given sentence.\n
You will get a sentence and you should rewrite it.\n
You should rewrite the sentence to be of this style '{type}'!\n
You should only answer with the your generated sentence, NOTHING ELSE!
\end{lstlisting}

\subsubsection{User-Prompt: Extend, Find Synonym \& Rewrite Sentence} \mbox{}
\begin{lstlisting}
{Sentence}
\end{lstlisting}

\subsubsection{User-Prompt: Find custom Sentence:} \mbox{}
\begin{lstlisting}
*sentence*: {sentence}\n
*prompt*: {prompt}    
\end{lstlisting}

\subsection{Experiment 1 - Texts}
\label{sec:appendix_texts_exp1}
\subsubsection{\Spread{}}
\paragraph{Extend the incomplete sentence (5 repetitions)}
Climate change refers to long-term shifts in temperatures and weather patterns. These shifts are increasingly driven by human activities such as the burning of fossil fuels, deforestation, and industrial processes, which
\paragraph{Extend the text by one sentence (3 repetitions) \& Extend the text by three sentences (3 repetitions)}
Climate change refers to long-term shifts in temperatures and weather patterns. These shifts may be natural, such as through variations in the solar cycle.

\subsubsection{\Pinch}
\paragraph{Remove the incomplete sentence (5 repetitions)}
The internet is a global network of interconnected computers that communicate using standardized protocols. This vast and ever-expanding network enables the rapid exchange of information across the world, facilitating everything from basic email communication to complex
\paragraph{Shorten the text by one sentence (3 repetitions) \& Shorten the text by three sentences (3 repetitions)}
The internet is a global network of interconnected computers that communicate using standardized protocols. It allows users to access and share information quickly and easily. The internet supports various services, including email, social media, and online shopping. It has transformed how we communicate, learn, and conduct business worldwide.

\subsubsection{Combinations}
\paragraph{Extend by two sentences, then
remove one \& Shorten by two sentences and then add one (once each)}
Photosynthesis is the process by which green plants, algae, and some bacteria convert light energy into chemical energy. They use sunlight, carbon dioxide, and water to produce glucose and oxygen. Chlorophyll, the green pigment in plants, captures sunlight for this process. Photosynthesis is essential for life on Earth, providing food and oxygen for most living organisms.

\subsection{Experiment 2 - Texts}
\label{sec:appendix_texts_exp2}
\subsubsection{ChatGPT Interface}
\lbparagraph{Repetition 1}
\noindent\textit{Instruction: Remove the irrelevant sentence.}

Tomatoes are one of the most popular vegetables grown in home gardens. Skateboards have been popular since the 1950s and are used in various sports competitions. Many people enjoy growing tomatoes because they are relatively easy to cultivate and can be used in a wide range of dishes.

\noindent\textit{Instruction: Notice the topic shift and extend.}

The process of growing tomatoes involves several key steps. Firstly, tomatoes require well-drained soil that is rich in organic matter. This helps the plants to thrive and produce a good yield. By following these steps, you can ensure a successful tomato harvest.
\lbparagraph{Repetition 2}
\noindent\textit{Instruction: Remove the irrelevant sentence.}

The art of photography has evolved significantly over the years. Bicycles have been a popular mode of transportation for over a century and are widely used for commuting and recreation. Many people find photography to be a rewarding hobby because it allows them to capture and preserve memories.

\noindent\textit{Instruction: Notice the topic shift and extend.}

To capture a great photograph, there are a few essential tips to keep in mind. Firstly, understanding the basics of lighting is crucial, as it can dramatically affect the mood and clarity of an image. By mastering these techniques, you can significantly improve the quality of your photos.

\lbparagraph{Repetition 3}
\noindent\textit{Instruction: Remove the irrelevant sentence.}

Reading books is a popular pastime that can enrich one's knowledge and imagination. Computers have revolutionized the way we work and communicate, making tasks easier and more efficient. Many people enjoy reading because it allows them to escape into different worlds and perspectives.

\noindent\textit{Instruction: Notice the topic shift and extend.}

To fully enjoy reading, it's important to choose books that match your interests and reading level. Firstly, selecting a quiet and comfortable place to read can enhance your concentration and enjoyment. By taking these steps, you can make reading a more fulfilling experience.

\subsubsection{Touch Gestures}
\lbparagraph{Repetition 1}
\noindent\textit{Instruction: Remove the irrelevant sentence.}

Cooking at home can be a rewarding experience that allows you to experiment with different ingredients and flavors. Automobiles have become an essential part of modern life, providing convenience and mobility. Many people find cooking to be a creative outlet that also promotes healthier eating habits.

\noindent\textit{Instruction: Notice the topic shift and extend.}

There are a few key tips to keep in mind when cooking. Firstly, it's important to use fresh ingredients, as they can significantly enhance the taste and nutritional value of your dishes. By following these tips, you can improve your cooking skills and enjoy better meals.
\lbparagraph{Repetition 2}
\noindent\textit{Instruction: Remove the irrelevant sentence.}

Gardening is a relaxing activity that allows you to connect with nature and grow your own plants. The invention of the airplane has made long-distance travel faster and more accessible. Many people enjoy gardening because it provides a sense of accomplishment and improves the beauty of their surroundings.

\noindent\textit{Instruction: Notice the topic shift and extend.}

Successful gardening requires some basic knowledge and attention to detail. Firstly, understanding the specific needs of your plants, such as sunlight and watering, is crucial for their growth. By adhering to these guidelines, you can create a thriving garden that brings you joy.

\lbparagraph{Repetition 3}
\noindent\textit{Instruction: Remove the irrelevant sentence.}

Exercise is an important aspect of maintaining a healthy lifestyle. Smartphones have changed the way we interact with the world, providing instant access to information and communication. Regular physical activity can help improve both mental and physical well-being.

\noindent\textit{Instruction: Notice the topic shift and extend.}

There are several factors to consider when establishing a workout routine. Firstly, it's essential to set realistic goals that align with your fitness level and interests. By doing so, you can create a sustainable exercise plan that keeps you motivated.

\subsection{Semi-Structured Interview - Short Stories}
\label{sec:short_stories}
We offered participants the beginning of creative short stories if they wished to explore our prototype during the semi-structured interview that concluded the study session. 
However, they were free to interact with the prototype in any way they wanted. 
The following short stories were created using ChatGPT 4o and individually reviewed for potential biases:

\paragraph{Time Travel}
In the year 3021, a young scientist named Finn discovered a device that could send him back in time. When he accidentally traveled to 1921, he had to figure out how to return without altering history. Along the way, he uncovered a hidden truth about his own family that changed everything.

\paragraph{Magical Adventure}
One day, Lena found an old compass in an abandoned antique shop, but instead of pointing north, it led her to a secret, enchanted forest. As she followed the compass, she encountered talking animals and an ancient, wise tree that entrusted her with an important task. With bravery and cleverness, Lena solved the forest's riddle and discovered a treasure far more valuable than gold.

\paragraph{Space Expedition}
Captain Mira and her crew landed on an unknown planet covered in glowing crystals that emitted a strange energy. As they explored the mysterious caves, they uncovered an ancient civilization trapped within the crystals. Mira faced a tough decision: should she destroy the crystals and free the beings inside, even if it meant risking their mission?

\paragraph{Enchanted Market}
At the annual Wonderland Market, little Timmy stumbled upon a stall selling wishes in bottles. Mesmerized by the shimmering elixirs, he bought a bottle that promised to make his wildest dreams come true. When he made his wish, a portal opened to a world beyond his imagination, full of adventure and danger.

\paragraph{Lost City}
Deep in the jungle, archaeologist Dr. Elena found an ancient map that revealed the way to a lost city of gold. With a team of explorers, she embarked on a perilous journey through rivers and over mountains, until they finally stood before the gates of the legendary city. But the city was not abandoned, and its mysterious inhabitants had their own plans for the intruders.

\begin{table*}[!h]
\centering
\footnotesize
\newcolumntype{L}{>{\raggedright\arraybackslash}X}
\newcolumntype{P}[1]{>{\raggedright\arraybackslash}p{#1}}
\renewcommand{\arraystretch}{1.4}
\setlength{\tabcolsep}{4pt}
\begin{tabularx}{\linewidth}{lP{2.75em}P{5.25em}P{22em}P{7em}L}
\toprule
    &
    \textbf{Section} &
    \textbf{Aspect}\newline and model &
    \textbf{Predictors}\newline Baseline (Exp. 1): \visnone\newline Baseline (Exp. 2): \modegpt &
    \textbf{Follow-up comparisons} &
    \textbf{Takeaways in words}\newline(only considering sig. results) \\ \midrule
1 &
    \ref{ssec:time}
    &
    Completion time (Exp. 1)\medskip\newline
    \textit{LMM on seconds}
    &
    \visbubble{} $\downarrow^*$ \newline 
    \deemph{(\lmmci{-2.62}{.76}{-4.11}{-1.13}{<.005})}\medskip\newline 
    \visline{} $\downarrow$ \newline 
    \deemph{(\lmmci{-.85}{.77}{-2.37}{.67}{.27})}\medskip\newline 
    &
    \visbubble{} vs \visline{}\newline \deemph{(\posthoc{-1.76}{<.05})}
    &
    People finished the tasks (in experiment 1) \secs{1.76} faster with \visbubble{} than with \visline{} and \secs{2.62} faster than without visual feedback.
    \\
    \midrule
2 &
    \ref{ssec:time}
    &
    Completion time (Exp. 2)\medskip\newline
    \textit{LMM on seconds}
    &
    \modeours{} $\downarrow^*$ \newline 
    \deemph{(\lmmci{-79.23}{10.81}{-100.34}{-57.21}{<.0001})}\medskip\newline 
    &
    &
    People finished the tasks (in experiment 2) \secs{79.23} faster with \modeours{}{} than with \modegpt{}.
    \\
    \midrule
3 &
    \ref{sec:perception_exp1}
    &
    Usability (Exp. 1)\medskip\newline
    \textit{LMM on SUS scores}
    &
    \visbubble{} $\uparrow^*$ \newline 
    \deemph{(\lmmci{22.37}{5.29}{11.85}{32.63}{<.001})}\medskip\newline 
    \visline{} $\uparrow^*$ \newline 
    \deemph{(\lmmci{13.80}{5.29}{3.28}{24.06}{<.05})}\medskip\newline 
    &
    \visbubble{} vs \visline{}\newline \deemph{(\posthoc{8.57}{=.11})}
    &
    The perceived usability (as measured with the SUS score) of \visbubble{} was higher than \visnone{} (by ca. 22 points). The score of \visline{} was also higher than that of \visnone{} (by ca. 14 points).
    \\
    \midrule
4 &
    \ref{sec:perception_exp2}
    &
    Usability (Exp. 2)\medskip\newline
    \textit{LMM on SUS scores}
    &
    \modeours{} $\uparrow^*$ \newline 
    \deemph{(\lmmci{28.46}{6.72}{15.05}{42.10}{<.001})}\medskip\newline  
    &
    &
    The perceived usability of \modeours{}{} was higher than that of \modegpt{} (by ca. 28 points).
    \\
    \midrule
5 &
    \ref{sec:perception_exp1}
    &
    Workload (Exp. 1)\medskip\newline
    \textit{LMM on NASA TLX scores}
    &
    \visbubble{} $\downarrow^*$ \newline 
    \deemph{(\lmmci{-.83}{.31}{-1.44}{-0.23}{<.05})}\medskip\newline 
    \visline{} $\downarrow^*$ \newline 
    \deemph{(\lmmci{-.66}{.31}{-1.26}{-0.05}{<.05})}\medskip\newline 
    &
    \visbubble{} vs \visline{}\newline \deemph{(\posthoc{-.18}{=.56})}
    &
    The perceived workload (as measured with the NASA TLX score) of \visbubble{} was lower than \visnone{} (by 0.83). The score of \visline{} was also lower than that of \visnone{} (by 0.66).
    \\
    \midrule
6 &
    \ref{sec:perception_exp2}
    &
    Workload (Exp. 2)\medskip\newline
    \textit{LMM on NASA TLX scores}
    &
    \modeours{} $\downarrow^*$ \newline 
    \deemph{(\lmmci{-1.07}{.32}{-1.74}{-.43}{<.01})}\medskip\newline  
    &
    &
    The perceived workload of \modeours{}{} was lower than that of \modegpt{} (by 1.07).
    \\
  \bottomrule
\end{tabularx}
\caption{Statistical tests for our analysis. Columns show link to the section, tested measure, predictors, follow-up comparisons, and a textual interpretation. We use arrows to highlight whether predictors increase ($\uparrow$) or decrease ($\downarrow$) the outcome and add an asterix  (*) if this is significant.}
\Description{Overview of significance tests with links to the section, tested measure, predictors, pairwise comparisons, and written interpretation. For each statistical test it describes the Section, Aspect and model, Predictors, Pairwise comparisons, and Takeaways in words (only considering sig. results).}
\label{tab:sig_tests}
\end{table*}

\subsection{Statistical Analyses}\label{sec:appendix_sigtest}

\cref{tab:sig_tests} shows our statistical analyses.
We analysed the data with linear mixed-effects models (LMMs), using R~\cite{R2020} with the \textit{lme4}~\cite{Bates2015} and \textit{lmerTest}~\cite{Kuznetsova2017} packages. Besides the fixed effects seen in the table, the models included random intercepts to account for individual differences between participants and between the tasks (texts). 
The follow-up analyses (pairwise comparisons) were conducted with the \textit{emmeans} package, using Bonferroni-Holm correction.
We report significance at p~<~0.05.

\subsection{\revision{Additional Figures}} \label{sec:appendix_figs}
\begin{figure*}[h!]
     \centering
     \begin{subfigure}[b]{0.49\textwidth}
        \centering
        \includegraphics[width=0.95\linewidth]{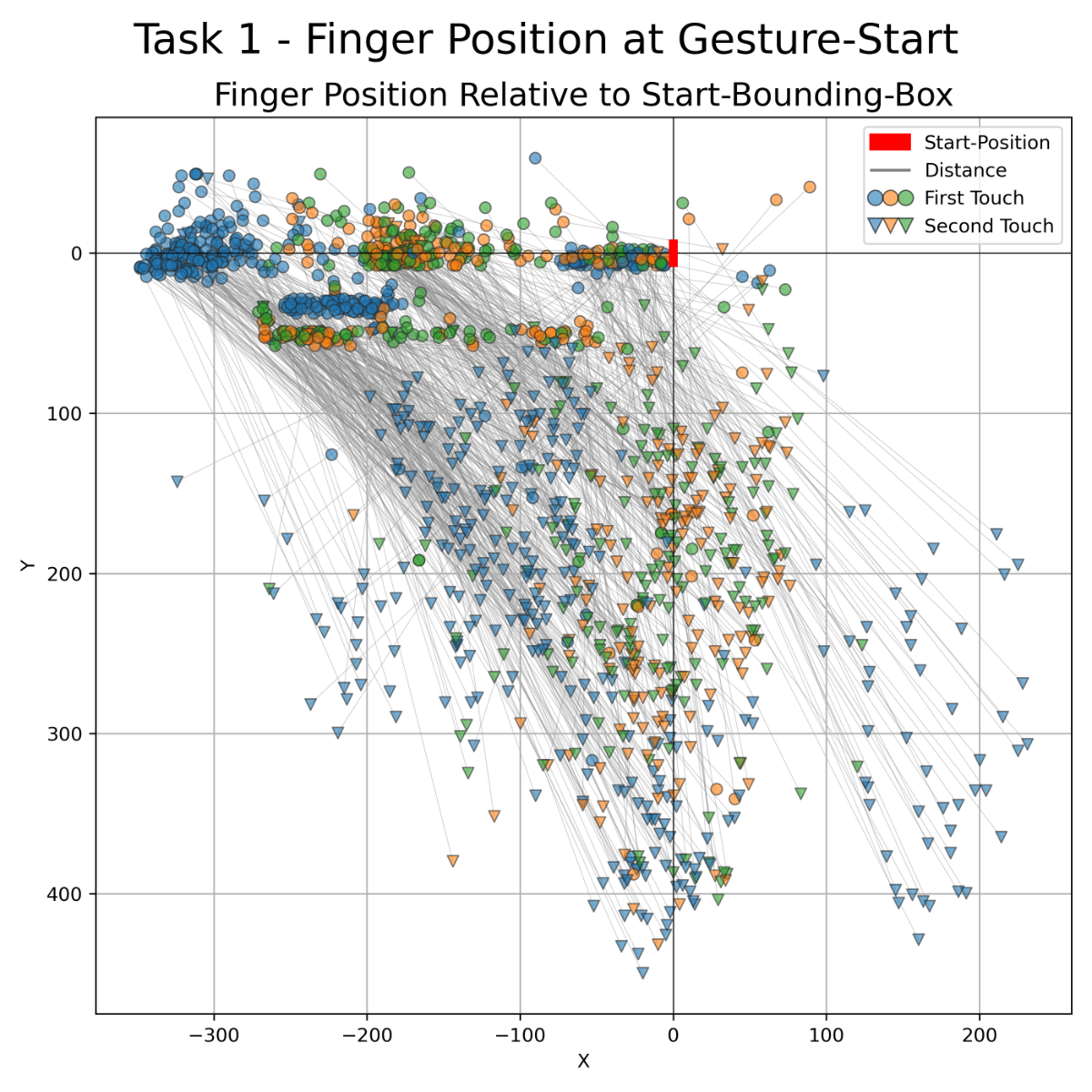}
        \caption{First and second touches in Experiment 1, when initiating a \spread{}.}
        \label{fig:first_touches_add}
     \end{subfigure}
     \hfill
     \begin{subfigure}[b]{0.49\textwidth}
        \centering
        \includegraphics[width=0.95\linewidth]{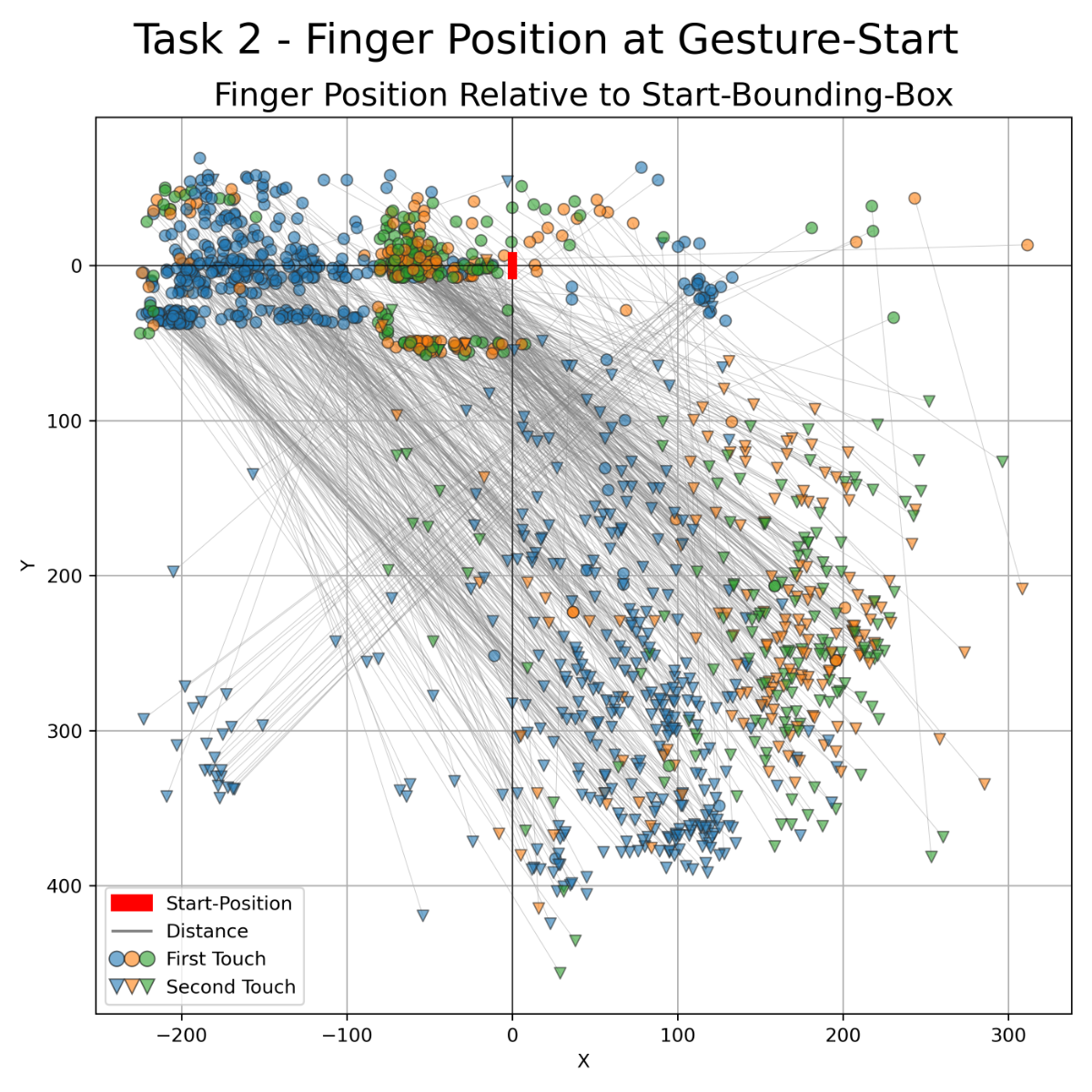}
        \caption{First and second touches for all participants in Experiment 1, when initiating a \pinch{} gesture.}
        \label{fig:first_touches_remove}
     \end{subfigure}
     \caption{Finger positions at gesture start for all participants in Experiment 1, when initiating a \spread{} (a), and a \pinch{} (b) gesture. In both subfigures, touches are plotted relative to the bounding box of the target area, with 'X' and 'Y' coordinates showing the spread of finger positions across participants. The colours indicate the starting text: blue -- incomplete sentence, orange -- one sentence, green -- three sentences.}
     \Description{This figure contains two scatter plots showing finger positions at the start of gestures for all participants in Task 1. Each plot visualises the initial finger touches for two distinct types of gestures: \spread{} and \pinch{}. Both plots use colour-coded markers and connecting lines to represent the trajectories of these finger movements. (a) The left-hand scatter plot shows the finger positions when participants initiated a \spread{} gesture. Each marker colour represents finger touches for different sub-tasks. The red vertical line indicates the central start position for the gestures. Thin grey lines connect the first and second touch points to illustrate the gesture's trajectory. (b) The right-hand scatter plot visualises the finger positions when initiating a \pinch{} gesture. Similar to (a), the first and second finger touches for each sub-task are indicated by different coloured markers. Grey lines represent the movement trajectory between the first and second touches. Both plots have an X and Y axis, with the X-axis showing the horizontal position of the touch relative to the centre, and the Y-axis showing the vertical position relative to the start bounding box. The overall layout helps to visualise the variation in finger positions and movement patterns across participants for both gestures.}
     \label{fig:first_touches}
\end{figure*}

\end{document}